\documentclass[twocolumn]{aastex6}

\shorttitle{Obliquity and Eccentricity Constraints}
\shortauthors{Stephen R. Kane \& Stephanie M. Torres}

\begin{document}

\title{Obliquity and Eccentricity Constraints For Terrestrial
  Exoplanets}

\author{
  Stephen R. Kane\altaffilmark{1,2},
  Stephanie M. Torres\altaffilmark{2}
}
\altaffiltext{1}{Department of Earth Sciences, University of
  California, Riverside, CA 92521, USA}
\altaffiltext{2}{Department of Physics \& Astronomy, San Francisco
  State University, 1600 Holloway Avenue, San Francisco, CA 94132,
  USA}
\email{skane@ucr.edu}


\begin{abstract}

Exoplanet discoveries over recent years have shown that terrestrial
planets are exceptionally common. Many of these planets are in compact
systems that result in complex orbital dynamics. A key step toward
determining the surface conditions of these planets is understanding
the latitudinally dependent flux incident at the top of the atmosphere
as a function of orbital phase. The two main properties of a planet
that influence the time-dependent nature of the flux are the obliquity
and orbital eccentricity of the planet. We derive the criterion for
which the flux variation due to obliquity is equivalent to the flux
variation due to orbital eccentricity. This equivalence is computed
for both the maximum and average flux scenarios, the latter of which
includes the effects of the diurnal cycle. We apply these calculations
to four known multi-planet systems (GJ~163, K2-3, Kepler-186, and
Proxima Centauri), where we constrain the eccentricity of terrestrial
planets using orbital dynamics considerations and model the effect of
obliquity on incident flux. We discuss the implications of these
simulations on climate models for terrestrial planets and outline
detectable signatures of planetary obliquity.

\end{abstract}



\section{Introduction}
\label{intro}

Exoplanetary science is rapidly requiring the need for
characterization techniques for terrestrial planets as their discovery
rate continues to increase. The {\it Kepler} mission has demonstrated
that planet frequency increases with smaller size
\citep{fre13,how13,pet13}, implying that the {\it Transiting Exoplanet
  Survey Satellite} will discover numerous examples of terrestrial
planets around bright host stars \citep{ric15,sul15}. Significant
attention is given to those planets that lie within the Habitable Zone
(HZ) of their host stars \citep{kas93,kop13,kop14}, although the HZ is
primarily a target selection tool for future atmospheric studies
\citep{kan12a}. In the meantime, General Circulation Models (GCMs) are
used to provide our best estimate of the surface conditions for
discovered HZ planets
\citep{wor10,wor11,lec13,wol13,yan13,wol14,yan14,lec15,kop16}.

A primary driving force in GCMs affecting surface conditions, climate
dynamics, and seasonal variations, is the instellation flux on the
planet \citep{kas15}. Two primary factors effect the variability of
the instellation flux: orbital eccentricity and obliquity. Tidal
effects can occasionally play a significant role for planets in
eccentric orbits and/or involved in planet-planet interactions
\citep{bar08,bar09} and may even push the planet into a runaway
greenhouse scenario \citep{bar13,dri15}. The effect of eccentricity on
planetary atmospheres and subsequent climate variations follows a
Keplerian pattern of long winters interrupted by brief periods of
``flash heating'' during periastron passage
\citep{wil02,dre10,kan12b,bol16}. The obliquity of a planet's
rotational axis undergoes short and long term oscillations due to
perturbations from other planetary bodies in the system
\citep{las86}. Fluctuations in planetary obliquity can have large
effects on climates \citep{wil97} and extreme obliquities can move the
outer edge of the HZ \citep{wil03,arm14,lin15}. Of the two primary
factors, orbital eccentricity is currently a far more accessible
measurable than obliquity. However, for systems in which we have
constraints on eccentricity, we can determine the range of obliquities
that drive the variation of instellation flux.

\begin{figure*}
  \begin{center}
    \includegraphics[angle=270,width=16.0cm]{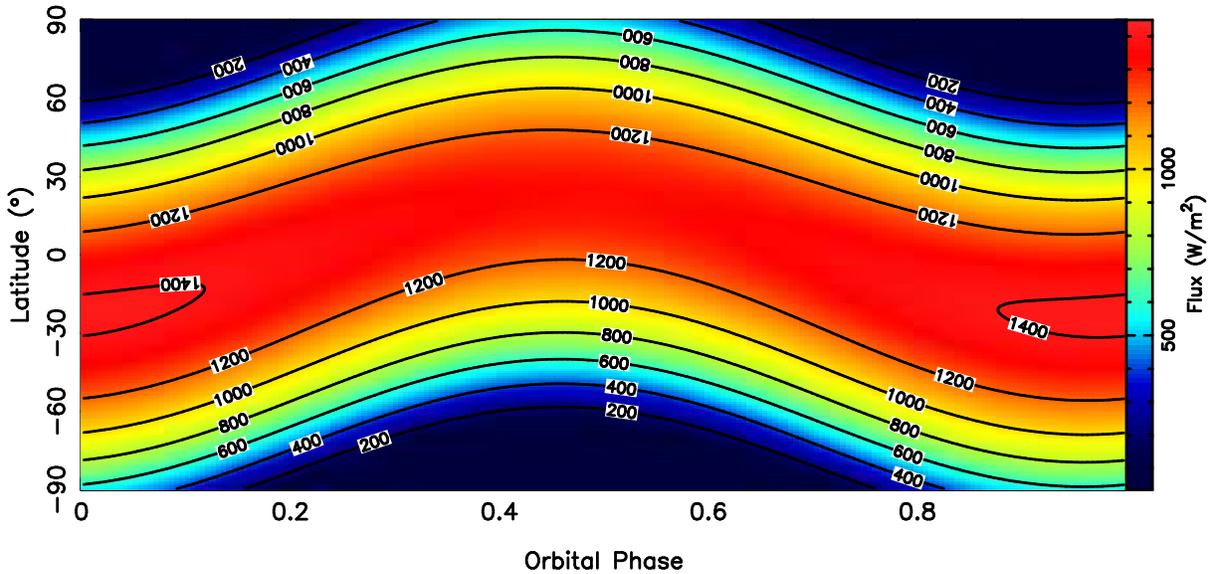}
  \end{center}
  \caption{A map of the maximum incident flux received by an Earth
    analog as a function of latitude and phase. The phase of $\phi =
    0.0$ corresponds to periastron passage of the planet.}
  \label{earth1}
\end{figure*}

Here, we describe the latitudinal flux incident on an exoplanet as a
function of obliquity, eccentricity, and orbital phase. We further
show how eccentricity constraints from radial velocity (RV)
measurements or dynamical constraints can be used to model
obliquity-dependent flux variations and locate regions where the
changes in flux due to obliquity are equivalent to those due to
eccentricity. In Section~\ref{flux} we formulate the time variable
flux and equate regions of flux change in obliquity and eccentricity
parameter space, both for maximum and average flux scenarios. In
Section~\ref{casestudies} we provide stability criteria for known
terrestrial planets in the GJ~163, K2-3, Kepler-186, and Proxima
Centauri systems and model their potential flux maps as a function of
obliquity. In Section~\ref{implications} we discuss the implications
of the flux variations for surface temperatures and atmospheric
conditions in so far as they effect habitability. We provide
concluding remarks and suggestions for future work in
Section~\ref{conclusions}.


\section{The Time Variable Flux}
\label{flux}

Here, we consider the orbital eccentricity and the obliquity of the
planetary rotation axis as sources of variable flux as a function of
latitude. For a given eccentricity, $e$, semi-major axis, $a$, and
star-planet separation, $r$, the maximum flux occurs at periastron, $r
= a(1-e)$, and the minimum flux occurs at apastron, $r = a(1+e)$. In
Section~\ref{maxflux}, we calculate maximum incident flux (when the
star crosses the local meridian) for a given latitude. This maximum
flux can be considered the instantaneous flux, or the latitudinal flux
received as a planet nears synchronous rotation (tidal locking). In
Section~\ref{aveflux}, we calculate the average the flux over the
diurnal cycle of the planet, which applies to planets whose rotation
period is significantly smaller than the orbital period.

\subsection{Maximum Flux Variation}
\label{maxflux}

The maximum flux at latitude $\beta$ is given by
\begin{eqnarray}
  F &=& \frac{L_\star}{4 \pi r^2} (\sin \delta \sin \beta + \cos
  \delta \cos \beta) \nonumber \\
  &=& \frac{L_\star}{4 \pi r^2} \cos | \beta - \delta |
  \label{flux1}
\end{eqnarray}
where $L_\star$ is the stellar luminosity. The solar declination,
$\delta$, is given by:
\begin{equation}
  \delta = \theta \cos [2 \pi (\phi - \Delta \phi)]
\end{equation}
for which $\phi$ is the orbital phase, $\Delta \phi$ is the offset in
phase between periastron and highest solar declination in the northern
hemisphere, and $\theta$ is the obliquity. For the Earth, $\Delta \phi
= 0.46$ and $\theta = 23.5\degr$. Figure~\ref{earth1} is an incident
flux map for an Earth--Sun analog as a function of latitude with
contours of constant flux throughout a complete orbital phase. The
phase of $\phi = 0.0$ corresponds to the planet's periastron passage.

The aim of the calculations here is to determine values of $e$ and
$\theta$ where the maximum change in flux during an orbit, $\Delta F$,
are equivalent at a given latitude, $\beta$. For the change in flux
due to $e$, we assume $\theta = 0\degr$, and likewise for the change
in flux due to $\theta$, we assume $e = 0$. For eccentricity, the
maximum change in flux is the difference in flux between periastron
and apastron:
\begin{eqnarray}
  \Delta F_e &=& \frac{L_\star}{4 \pi a^2 (1-e)} \cos \beta -
  \frac{L_\star}{4 \pi a^2 (1+e)} \cos \beta \nonumber \\
  &=& \frac{L_\star}{4 \pi a^2} \cos \beta \left( \frac{1}{1-e} -
  \frac{1}{1+e} \right) \nonumber \\
  &=& \frac{L_\star}{2 \pi a^2} \frac{e}{(1-e^2)} \cos \beta
  \label{ecc1}
\end{eqnarray}
For obliquity, the maximum change in flux occurs amidst the
difference between the minimum and maximum solar declination. When
$\theta \leq 45\degr$ and $\theta \leq \beta \leq 90\degr - \theta$,
this can be expressed as:
\begin{eqnarray}
  \Delta F_\theta &=& \frac{L_\star}{4 \pi a^2} \cos (\beta-\theta) -
  \frac{L_\star}{4 \pi a^2} \cos (\beta+\theta) \nonumber \\
  &=& \frac{L_\star}{4 \pi a^2} [\cos (\beta-\theta) - \cos
    (\beta+\theta)] \nonumber \\
  &=& \frac{L_\star}{2 \pi a^2} \sin \beta \sin \theta
  \label{ob1}
\end{eqnarray}
For $\beta < \theta$, the following applies:
\begin{equation}
  \Delta F_\theta = \frac{L_\star}{4 \pi a^2} [1 - \cos (\beta+\theta)]
  \label{ob2}
\end{equation}
and for $\beta > 90\degr - \theta$, the following applies:
\begin{equation}
  \Delta F_\theta = \frac{L_\star}{4 \pi a^2} \cos (\beta-\theta)
  \label{ob3}
\end{equation}

\begin{figure}
  \includegraphics[angle=270,width=8.2cm]{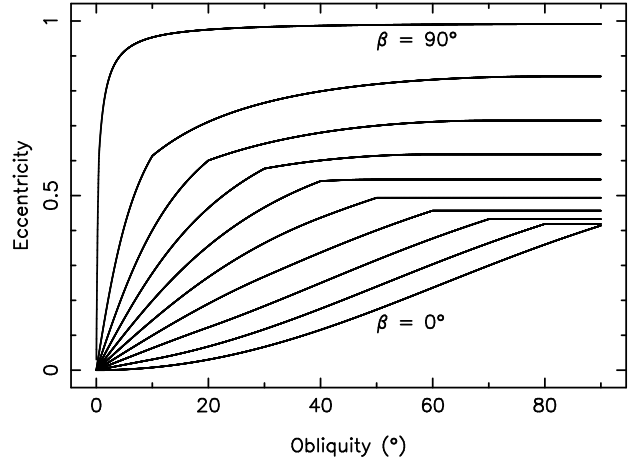}
  \caption{Orbital eccentricity as a function of the obliquity of the
    rotational axis. The lines of constant latitude represent
    equivalence of flux variation received ($\Delta F_e = \Delta
    F_\theta$) during one complete orbital period.}
  \label{fluxvar1}
\end{figure}

Thus, the maximum flux changes due to eccentricity and obliquity are
equivalent where $\Delta F_e = \Delta F_\theta$. Solving for obliquity
in the regime $\theta \leq \beta \leq 90\degr - \theta$, we combine
Equations \ref{ecc1} and \ref{ob1}:
\begin{eqnarray}
  \frac{L_\star}{2 \pi a^2} \frac{e}{(1-e^2)} \cos \beta &=&
  \frac{L_\star}{2 \pi a^2} \sin \beta \sin \theta \nonumber \\
  \frac{e}{(1-e^2)} &=& \tan \beta \sin \theta \nonumber \\
  \theta &=& \arcsin \left[ \frac{e}{(1-e^2) \tan \beta} \right]
  \nonumber
\end{eqnarray}
For $\beta < \theta$, we combine Equations \ref{ecc1} and \ref{ob2}:
\begin{eqnarray}
  \frac{L_\star}{2 \pi a^2} \frac{e}{(1-e^2)} \cos \beta &=&
  \frac{L_\star}{4 \pi a^2} [1 - \cos (\beta+\theta)] \nonumber \\
  \frac{e}{(1-e^2)} &=& \frac{1 - \cos (\beta+\theta)}{2 \cos \beta}
  \nonumber \\
  \theta &=& \arccos \left[ 1 - 2 \cos \beta \frac{e}{(1-e^2)} \right]
  - \beta \nonumber
\end{eqnarray}
For $\beta > 90\degr - \theta$, we combine Equations \ref{ecc1} and
\ref{ob3}:
\begin{eqnarray}
  \frac{L_\star}{2 \pi a^2} \frac{e}{(1-e^2)} \cos \beta &=&
  \frac{L_\star}{4 \pi a^2} \cos (\beta-\theta) \nonumber \\
  \frac{e}{(1-e^2)} &=& \frac{\cos (\beta-\theta)}{2 \cos \beta}
  \nonumber \\
  \theta &=& \beta - \arccos \left[ 2 \cos \beta \frac{e}{(1-e^2)}
    \right] \nonumber
\end{eqnarray}
Solving for eccentricity results in:
\begin{equation}
  e = \frac{\sqrt{1+f(\theta,\beta)^2} - 1}{f(\theta,\beta)}
  \label{eccfinal}
\end{equation}
where the function $f(\theta,\beta)$ for $\theta \leq 45\degr$ is
given by:
\begin{equation}
  f(\theta,\beta) = \left\{
  \begin{array}{ll}
    2 \tan \beta \sin \theta & \mbox{for $\theta \leq \beta \leq
      90\degr - \theta$}\\
    \frac{1 - \cos (\beta+\theta)}{\cos \beta} & \mbox{for $\beta <
      \theta$}\\
    \frac{\cos (\beta-\theta)}{\cos \beta} & \mbox{for $\beta >
      90\degr - \theta$}
  \end{array} \right.
  \label{obfinal}
\end{equation}
Equations \ref{eccfinal} and \ref{obfinal} allow the calculation of
orbital eccentricities for which the total change in incident flux is
the same as obliquities with $\theta \leq 45\degr$. Using the same
methodology for $\theta > 45\degr$, the function $f(\theta,\beta)$ is
given by:
\begin{equation}
  f(\theta,\beta) = \left\{
  \begin{array}{ll}
    \frac{1}{\cos \beta} & \mbox{for $\theta \geq \beta \geq
      90\degr - \theta$}\\
    \frac{1 - \cos (\beta+\theta)}{\cos \beta} & \mbox{for $\beta <
      90\degr - \theta$}\\
    \frac{\cos (\beta-\theta)}{\cos \beta} & \mbox{for $\beta >
      \theta$}
  \end{array} \right.
\end{equation}
Shown in Figure~\ref{fluxvar1} are the locations of eccentricity and
obliquity where the flux variation during a complete orbital phase are
equivalent to each other. We plot this for latitudes ranging from
$\beta = 0\degr$ to $\beta = 90\degr$ in steps of $10\degr$. At
latitudes close to the poles, the variation between winter and summer
incident flux increases as the pole is tilted toward the ecliptic
plane. The minimum flux for an eccentric orbit $e < 1$ will never
reach zero, even at apastron. Thus, an obliquity of $\theta = 90\degr$
approaches a boundary condition where the flux difference is
equivalent to that of a hyperbolic orbit.

\begin{figure*}
  \begin{center}
    \includegraphics[angle=270,width=16.0cm]{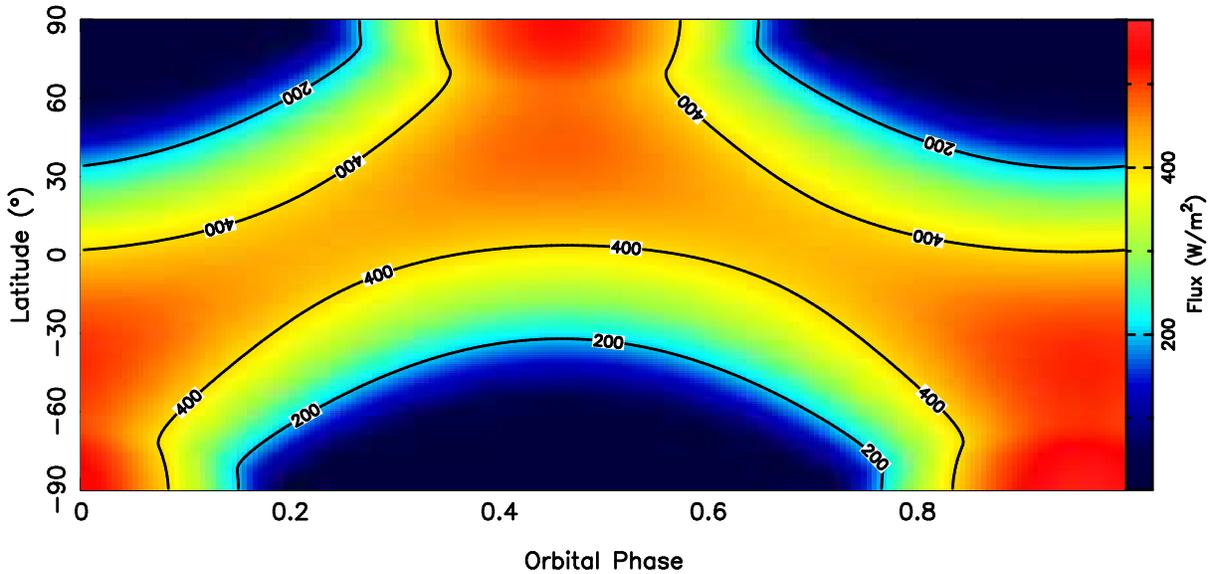}
  \end{center}
  \caption{A map of the incident flux received by an Earth analog,
    averaged over the diurnal cycle, as a function of latitude and
    phase. The phase of $\phi = 0.0$ corresponds to periastron passage
    of the planet.}
  \label{earth2}
\end{figure*}

There is a difference that should be noted for the change in flux due
to eccentricity and obliquity. Although the flux variation due to
obliquity at a given latitude varies sinusoidally, the flux variation
due to eccentricity varies based on the star-planet separation
produced by a Keplerian orbit. Therefore, though the total change in
flux is the same, the rate at which the flux varies between minimum
and maximum are different for the eccentricity and obliquity
scenarios, likely resulting in a different atmospheric response over
the orbital phase time scale.


\subsection{Diurnal Cycle Effects}
\label{aveflux}

For planets where the rotation period is significantly less than the
orbital period, the average incident flux as a function of latitude
may be used. For this purpose, Equation~\ref{flux1} is modified as
follows
\begin{equation}
  F = \frac{L_\star}{4 \pi r^2} (\sin \delta \sin \beta + \cos \delta
  \cos \beta \cos h)
  \label{flux2}
\end{equation}
where $h$ is the hour angle of the star with respect to the local
meridian. The fraction of planetary rotation period that experiences
daylight for a given latitude is
\begin{equation}
  \Delta t_{dl} = \frac{2 \arccos(-\tan \delta \tan \beta)}{360\degr}
  \label{daylight}
\end{equation}
For obliquities of $\theta > 0\degr$, there will be latitudes that
experience constant day/night during the course of an orbital
period. These situations are defined by the criteria that if $\beta +
\delta > 90\degr$ or $\beta + \delta < -90\degr$ then $\Delta t_{dl} =
1.0$, and if $\beta - \delta > 90\degr$ or $\beta - \delta < -90\degr$
then $\Delta t_{dl} = 0.0$. The average flux at a given latitude can
then be calculated by accounting for the change in flux as a function
of $h$ and the fractional daylight time. Figure~\ref{earth2} is an
incident flux map averaged over the diurnal cycle for an Earth--Sun
analog as a function of latitude. The comparison with
Figure~\ref{earth1} shows the impact of including the effect of
constant daylight periods on the polar regions.

\begin{figure}
  \includegraphics[angle=270,width=8.2cm]{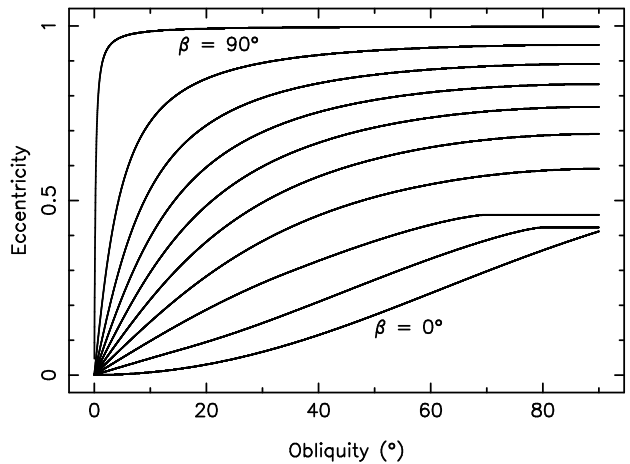}
  \caption{As for Figure~\ref{fluxvar1}, this figure shows the orbital
    eccentricity as a function of the obliquity of the rotational
    axis, but now includes the effect of the diurnal cycle. The lines
    of constant latitude represent equivalence of flux variation
    received ($\Delta F_e = \Delta F_\theta$) during one complete
    orbital period.}
  \label{fluxvar2}
\end{figure}

As for Section~\ref{maxflux}, we now calculate the values of $e$ and
$\theta$ where the change in the average flux during an orbit, $\Delta
F$, are equivalent at a given latitude, $\beta$. For the eccentricity
case with $\theta = 0\degr$, the average flux is equivalent to the
amplitude shown in Equation~\ref{ecc1} multiplied by the average of a
sine function including the effect of a day/night cycle. This leads to
an additional $1/\pi$ factor, as follows
\begin{equation}
  \Delta F_e = \frac{L_\star}{2 \pi^2 a^2} \frac{e}{(1-e^2)} \cos
  \beta
  \label{ecc2}
\end{equation}
For obliquity, the introduction of the hour angle and fraction
daylight in Equations \ref{flux2} and \ref{daylight} produce a
non-trivial calculation of $\Delta F_\theta$ for various obliquity and
latitude ranges. We solve this by numerically calculating regions
where $\Delta F_e = \Delta F_\theta$. The result of these calculations
are summarized in Figure~\ref{fluxvar2} where, as for
Figure~\ref{fluxvar1}, we plot lines of constant latitude from $\beta
= 0\degr$ to $\beta = 90\degr$ in steps of $10\degr$. The main effect
of including the diurnal cycle is to smooth the relationship with
eccentricity due to the averaging of the flux received for a given
latitude. Additionally, the diurnal cycle increases the equivalent
eccentricity at high latitudes, as the change in average flux is
larger than for the maximum flux case described in
Section~\ref{maxflux}. The combination of the two factors,
eccentricity and obliquity, are investigated for specific planets in
the case studies that follow.


\section{Case Studies}
\label{casestudies}

Here, we apply eccentricity constraints through stability
considerations to four of the known exoplanets: GJ~163~c, K2-3~d,
Kepler-186~f, and Proxima Centauri b. These are then used to determine
latitudinal flux maps of the planets as a function of orbital phase
for fixed obliquities, including diurnal effects. The four exoplanets
were carefully chosen from the known terrestrial exoplanets
considering their proximity to the HZ and the diversity of the system
architectures. System parameters were extracted from the NASA
Exoplanet Archive \citep{ake13} and relevant publications (see
Table~\ref{planets}).

\begin{deluxetable*}{lccccc}
  \tablecolumns{5}
  \tablewidth{0pc}
  \tablecaption{\label{planets} Stellar and Planetary Parameters}
  \tablehead{
    \colhead{Parameter} &
    \colhead{GJ~163~c$\,^{a}$} &
    \colhead{K2-3~d$\,^{b}$} &
    \colhead{Kepler-186~f$\,^{c}$} &
    \colhead{Proxima Centauri~b$\,^{d}$}
  }
  \startdata
\noalign{\vskip -3pt}
\sidehead{Star}
~~~~Spectral Type         & M3.5 V           & M0.0 V         & M1 V             & M5.5 V \\
~~~~$V$                   & $11.811\pm0.012$     & $12.17\pm0.01$       & 15.65            & 11.13 \\
~~~~Distance (pc)         & $15.0\pm0.4$         & $45\pm3$             & $151\pm18$       & 1.295 \\
~~~~$T_\mathrm{eff}$ (K)  & $3500\pm100$         & $3896\pm189$         & $3788\pm54$      & $3050\pm100$ \\
~~~~$M_\star$ ($M_\odot$) & $0.40\pm0.02$        & $0.60\pm0.09$        & $0.478\pm0.055$  & $0.120\pm0.015$ \\
~~~~$R_\star$ ($R_\odot$) & --                   & $0.56\pm0.07$        & $0.472\pm0.052$  & $0.141\pm0.021$ \\
~~~~$L_\star$ ($L_\odot$) & $0.0196\pm0.001$     & $0.065\,^{e}$        & $0.0412\pm0.069$ & $0.00155\pm0.00006$ \\
~~~~CHZ (AU)              & 0.145--0.282$\,^{e}$ & 0.262--0.500$\,^{e}$ & 0.21--0.40$\,^{e}$ & 0.041--0.081$\,^{e}$ \\
~~~~OHZ (AU)              & 0.115--0.297$\,^{e}$ & 0.207--0.527$\,^{e}$ & 0.17--0.42$\,^{e}$ & 0.032--0.086$\,^{e}$ \\
\sidehead{Planet}
~~~~$P$ (days)         & $25.63\pm0.03$    & $44.5631^{+0.0063}_{-0.0055}$ & $129.9459\pm0.0012$    & $11.186\pm0.002$ \\
~~~~$e$                & $0.099\pm0.086$   & $<0.162\,^{e}$                & $<0.628\,^{e}$         & $<0.35$ \\
~~~~$\omega$ ($\degr$) & $227\pm80$        & --                            & --                     & 310 \\
~~~~$M_p$ ($M_\oplus$) & $6.8\pm0.9$       & $3.97\,^{e}$                  & $1.54\,^{e}$           & $1.27^{+0.19}_{-0.17}$ \\
~~~~$R_p$ ($R_\oplus$) & --                & $1.52^{+0.21}_{-0.20}$        & $1.11^{+0.14}_{-0.13}$ & -- \\
~~~~$a$ (AU)           & $0.1254\pm0.0001$ & $0.2076^{+0.0098}_{-0.0108}$  & $0.356\pm0.048$        & $0.0485^{+0.0051}_{-0.0041}$ \\
~~~~$R_H$ (AU)         & --                & $0.004\,^{e}$                 & $0.005\,^{e}$          & --
  \enddata
  \tablenotetext{a}{\citet{bon13,tuo13}}
  \tablenotetext{b}{\citet{cro15}}
  \tablenotetext{c}{\citet{qui14}}
  \tablenotetext{d}{\citet{ang16}}
  \tablenotetext{e}{Calculated in this work.}
\end{deluxetable*}


\subsection{Stability Criteria}
\label{stability}

Of the four systems considered here, two were discoveries using the RV
technique (GJ~163 and Proxima Centauri). The planets in these systems
have measurements and subsequent constraints placed upon their orbital
eccentricities from the Keplerian fit to the RV data. The remaining
two systems, K2-3 and Kepler-186, were detected using the transit
method with scant RV data obtained. These two systems thus have
limited information available for the planetary orbital
eccentricities. Observations of compact Kepler systems indicate that
such planets are likely to be in circular orbits
\citep{kan12c,van15}. However, here we use stability considerations to
determine the maximum eccentricities allowed for planets in those
systems.

\begin{figure*}
  \begin{center}
    \begin{tabular}{cc}
      \includegraphics[angle=270,width=8.2cm]{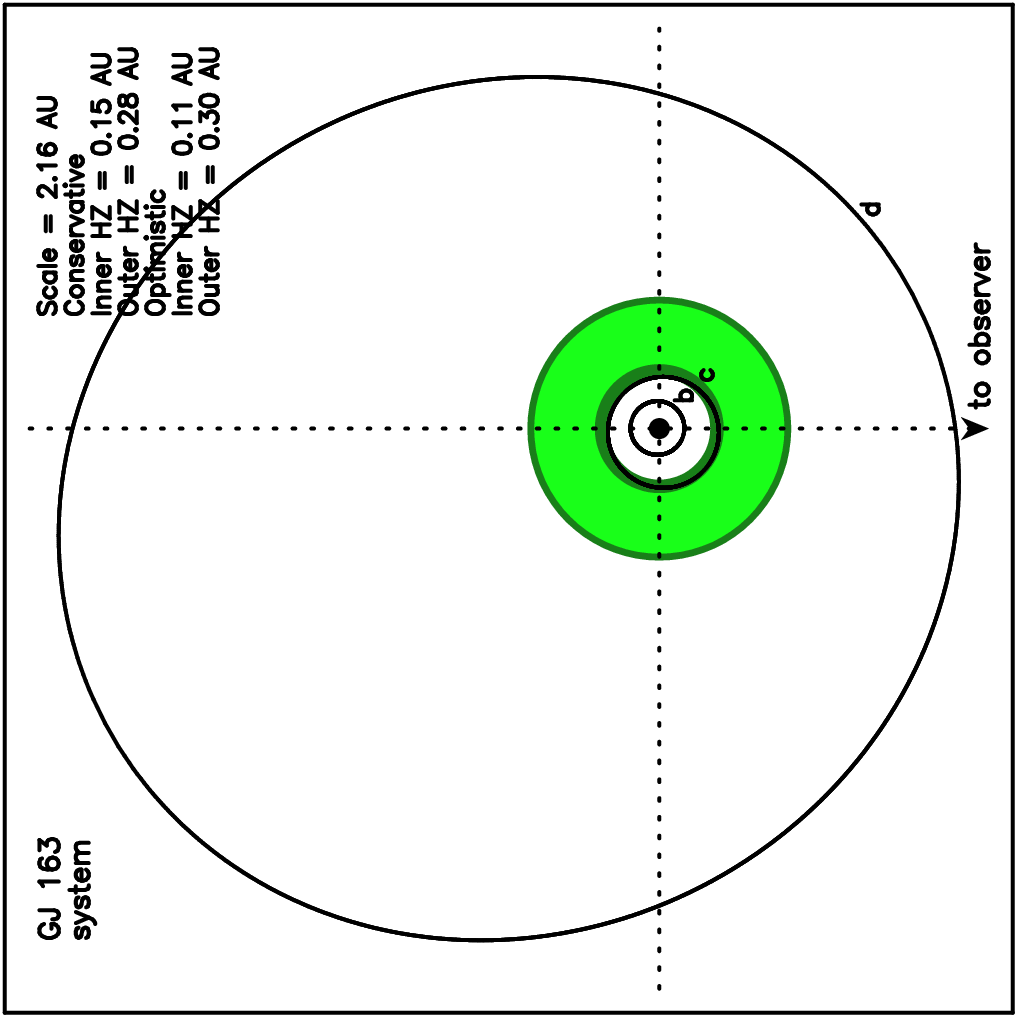} &
      \includegraphics[angle=270,width=8.2cm]{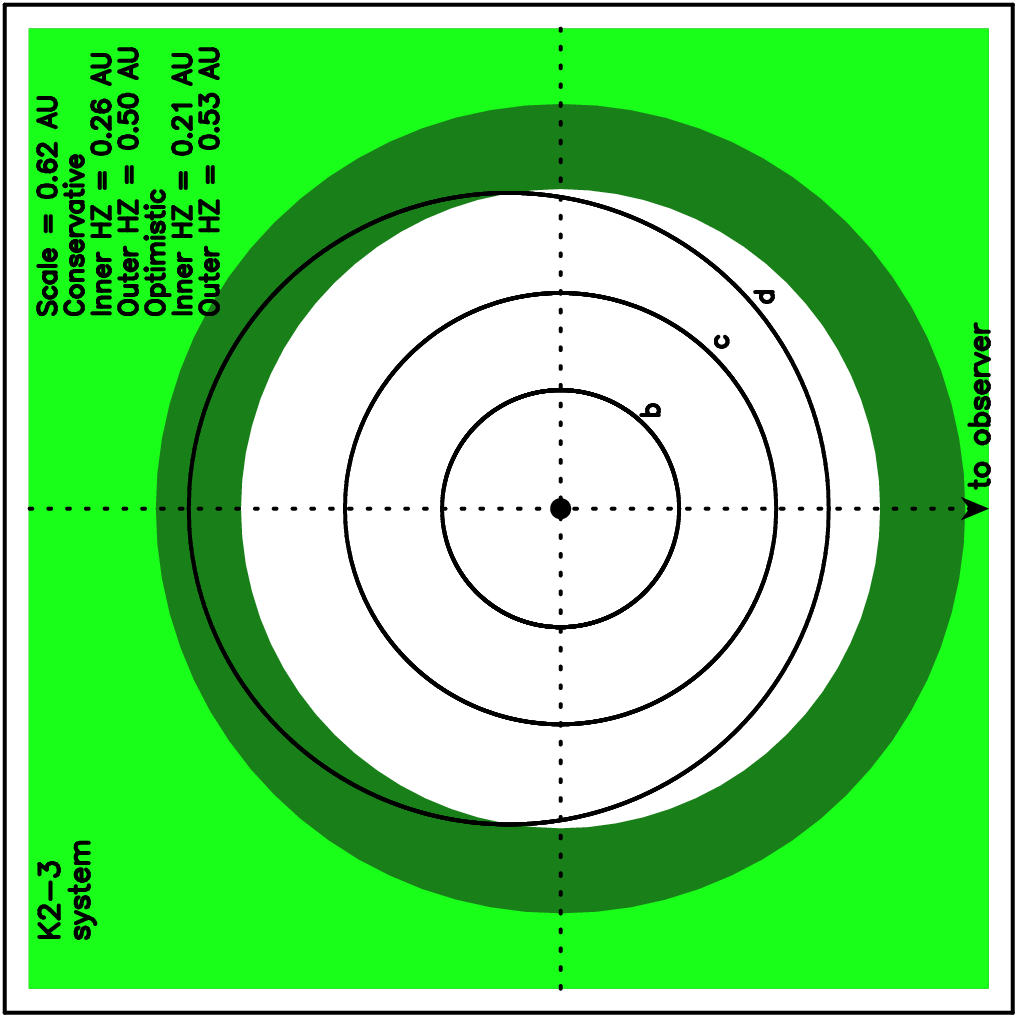} \\
      \includegraphics[angle=270,width=8.2cm]{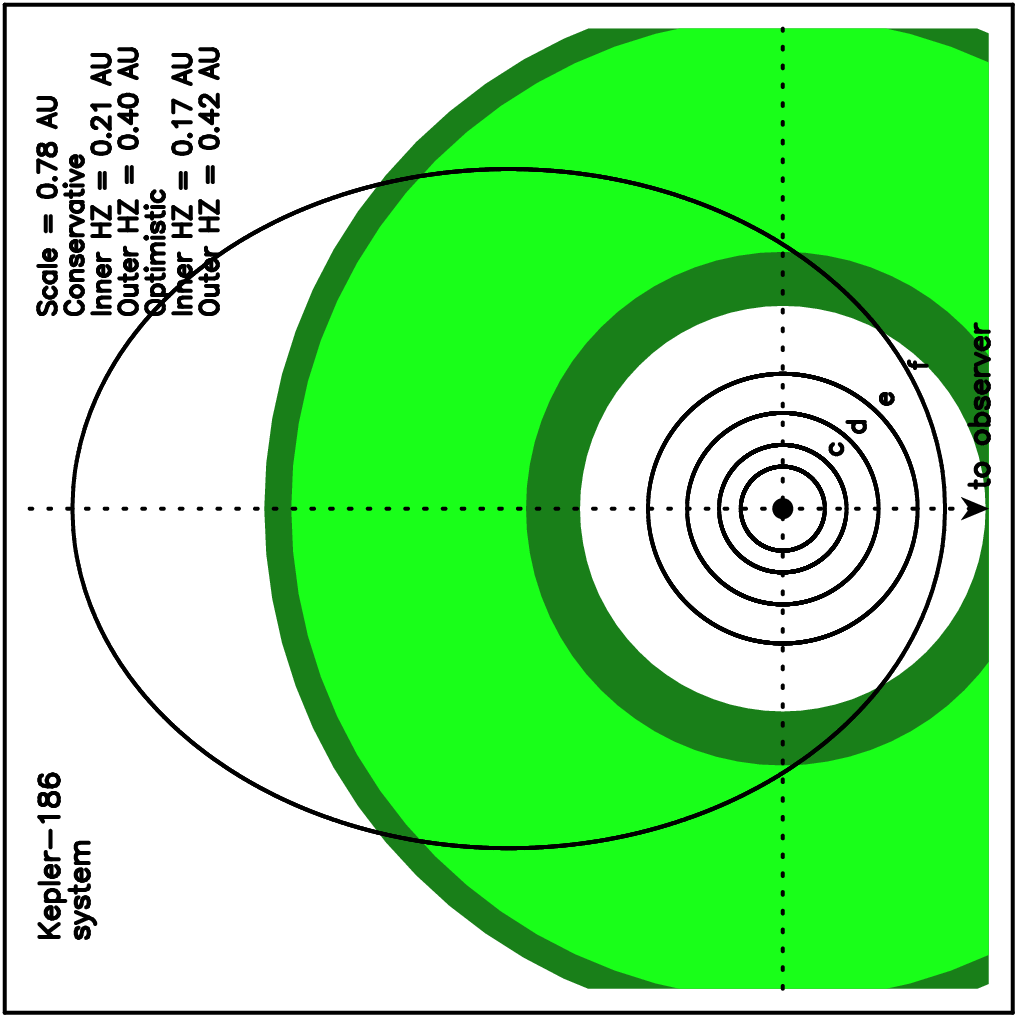} &
      \includegraphics[angle=270,width=8.2cm]{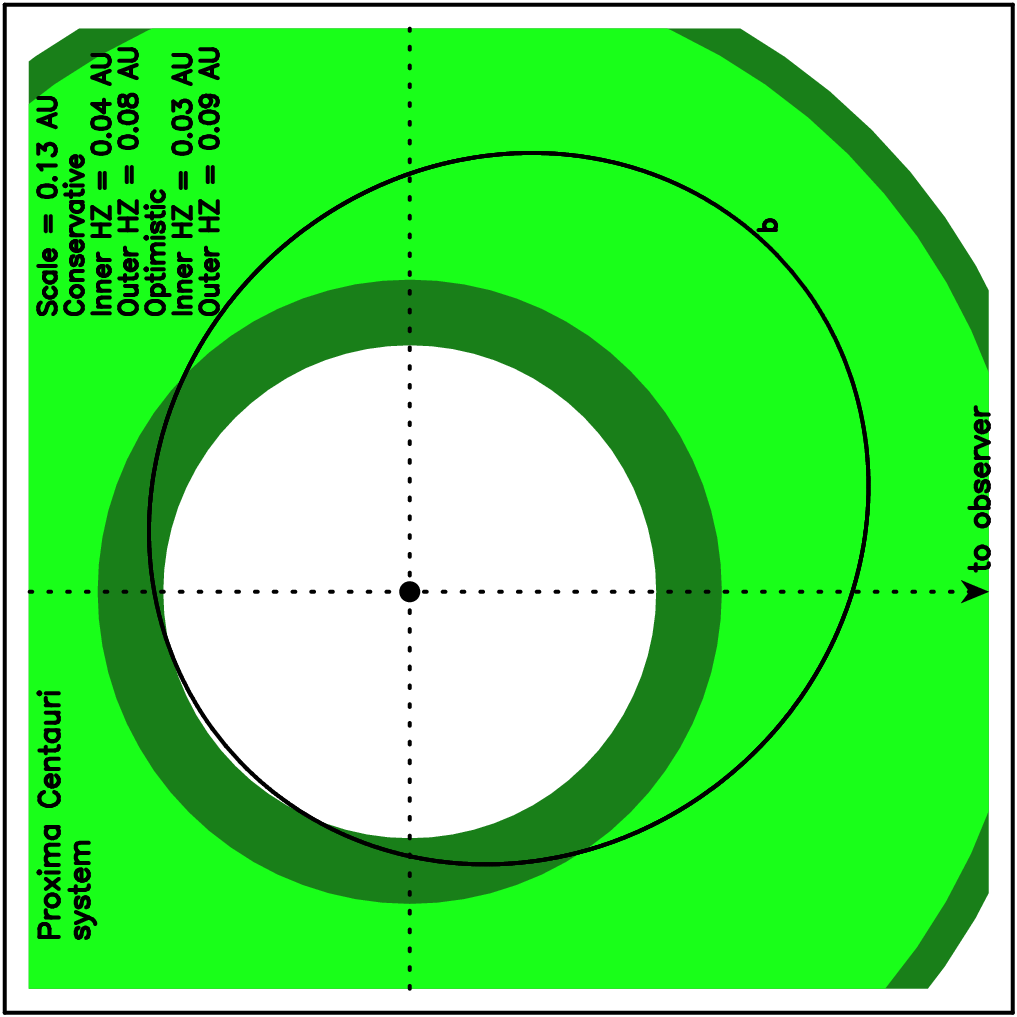}
    \end{tabular}
  \end{center}
  \caption{Top-down views of the GJ-163 (top-left), K2-3 (top-right),
    Kepler-186 (bottom-left), and Proxima Centauri (bottom-right)
    systems. The orbits of the planets are shown overlaid on the
    conservative (light-green) and optimistic (dark-green) HZ
    regions. The size along a panel side (scale) is indicated in the
    top-left corner of each panel. For K2-3~d and Kepler-186~f, the
    orbital eccentricities adopted are the maximum allowed from
    stability arguments (see Table~\ref{planets} and
    Section~\ref{stability}).}
  \label{hzfig}
\end{figure*}

We use a similar methodology for orbital stability to that used by
\citet{cro15} and \citet{sin16}. The masses of the transiting planets
($M_p$) were calculated using the mass-radius relationships of
\citet{wei14}. For two-planet systems, a criterion for stability was
numerically estimated by \citet{gla93}, requiring that the separation
of the planets exceed about 3.5 mutual Hill radii ($R_{H,M_p}$), given
by
\begin{equation}
  R_{H,M_p} = \left[ \frac{M_{p,in} + M_{p,out}}{3 M_\star}]
  \right]^{\frac{1}{3}} \frac{(a_{in} + a_{out})}{2}
    \label{hillradius}
\end{equation}
where $M_\star$ is the mass of the host star and the ``in/out''
subscripts refer to the inner and outer planets in the system. For
multi-planet systems, a long-term stability criterion established by
\citet{smi09} requires that $\Delta > 9$ for adjacent planets where
$\Delta = (a_{out} - a_{in})/R_H$. For three adjacent planets, the
criterion becomes $\Delta_{in} + \Delta_{out} > 18$, where
$\Delta_{in}$ and $\Delta_{out}$ are the $\Delta$ calculations for the
inner and outer adjacent planet pairs respectively. By modifying
Equation~\ref{hillradius} with a $(1 - e)$ multiplicative factor to
account for eccentricity, we are able to determine eccentricities that
satisfy the above stability criteria. The results of these
calculations for individual systems are described in the sections
specific to those systems below.


\subsection{Habitable Zone}
\label{hz}

To calculate the HZ boundaries of the four planetary systems studied
here, we use the methodology described by \citet{kop13,kop14}. There
are two inner and two outer boundaries calculated, the extent of which
depend on assumptions regarding how long Venus and Mars were able to
retain liquid water at their surfaces. These are referred to as the
Conservative Habitable Zone (CHZ) and the Optimistic Habitable Zone
(OHZ), for which a detailed description can be found in \citet{kop13}.
Our calculations for the CHZ and OHZ boundaries for each of the
systems are shown in Table~\ref{planets}.

Figure~\ref{hzfig} shows a top-down view of the systems, including the
planetary orbits and the CHZ (light-green) and OHZ (dark-green)
regions. The size along a panel side (scale) in the figure is
indicated in the top-right corner of each panel. The parameters used
to plot the planetary orbits are those from Table~\ref{planets} and
the associated references. For K2-3~d and Kepler-186~f, we have used
the maximum eccentricities for those planets using the calculations of
Section~\ref{stability} and described further in Sections \ref{k23}
and \ref{kepler186}. The percentage of a complete orbital period spent
within the OHZ for each of the four planets are 86\% (GJ~163~c), 56\%
(K2-3~d), 33\% (Kepler-186~f), and 94\% (Proxima Centauri b).


\subsection{GJ 163}
\label{gj163}

The known planets orbiting the low-mass star GJ~163 were discovered by
\citet{bon13}. Their analysis of the RV data indicated the presence of
five periodic signals, two of which were attributed to possible
stellar activity sources. The three-planet solution includes a
6.8~$M_oplus$ planet (planet c) in a $\sim$25~day period orbit. We
adopt this three-planet solution and use the stellar parameters of
\citet{bon13} and \citet{tuo13}, as shown in Table~\ref{planets}.

The Keplerian orbit of planet c reveals an orbital eccentricity of $e
\sim 0.1$ which we utilize in our models. According to
Figure~\ref{fluxvar1}, an obliquity of $\theta = 16\degr$ produces an
equivalent flux variation to that produced by the $e = 0.1$
eccentricity at a latitude of $\beta = 20\degr$. For a circular orbit
($e = 0.0$), the maximum flux received by planet c would be
1705~W\,m$^{-2}$ (1.25~$F_\oplus$). Using the measured eccentricity,
the maximum flux (during periastron passage) is 2100~W\,m$^{-2}$
(1.54~$F_\oplus$).

\begin{figure*}
  \begin{center}
    \includegraphics[angle=270,width=16.0cm]{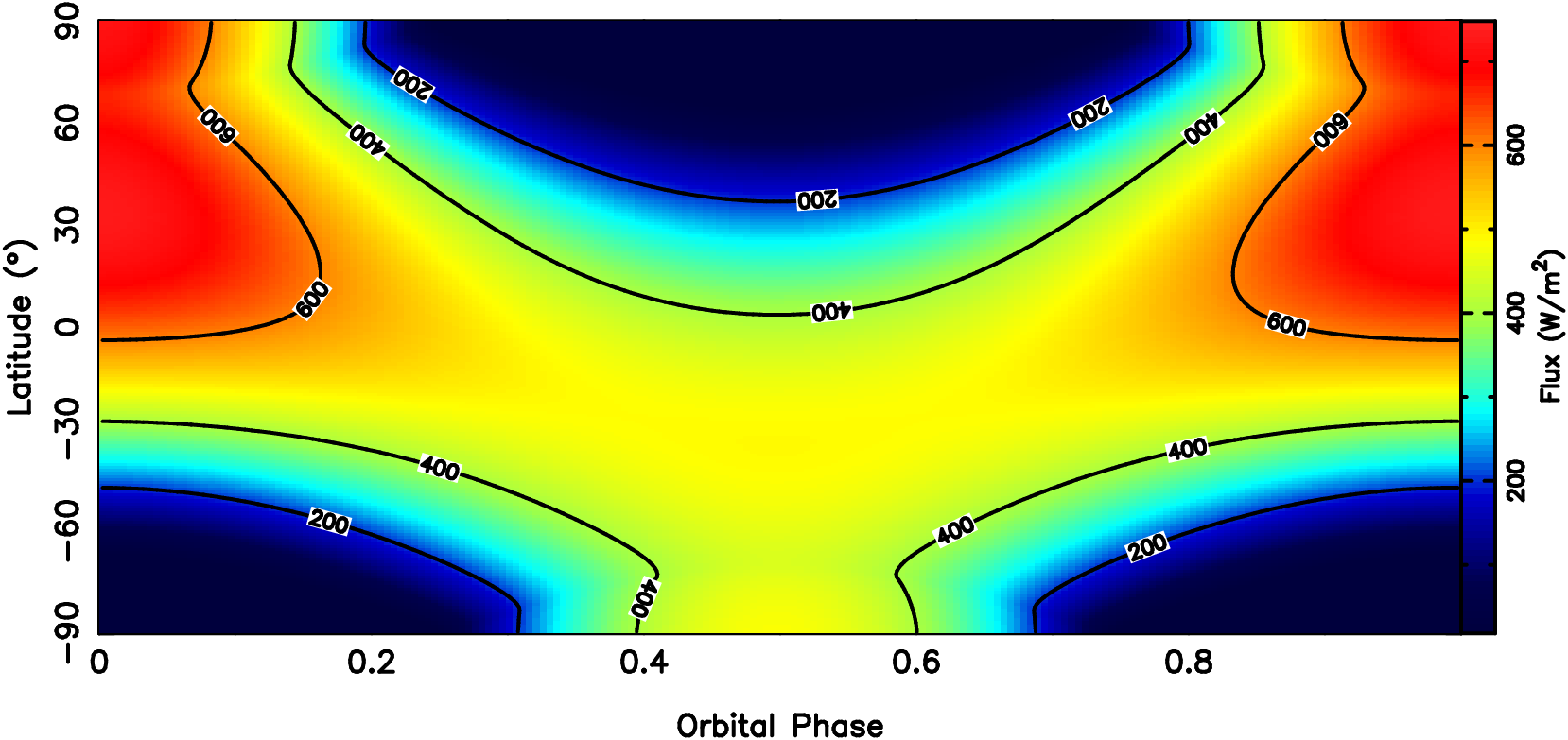} \\
    \includegraphics[angle=270,width=16.0cm]{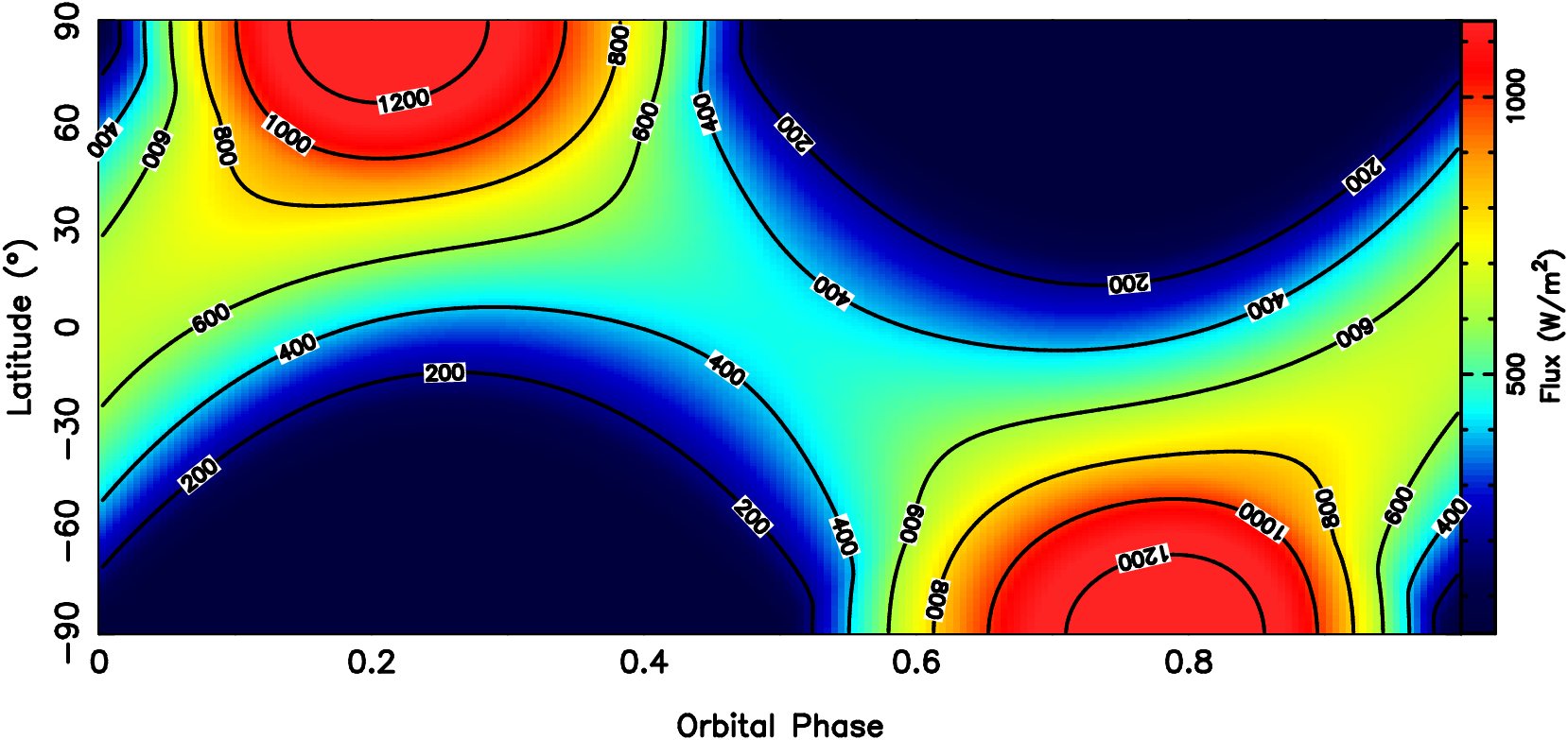} \\
    \includegraphics[angle=270,width=16.0cm]{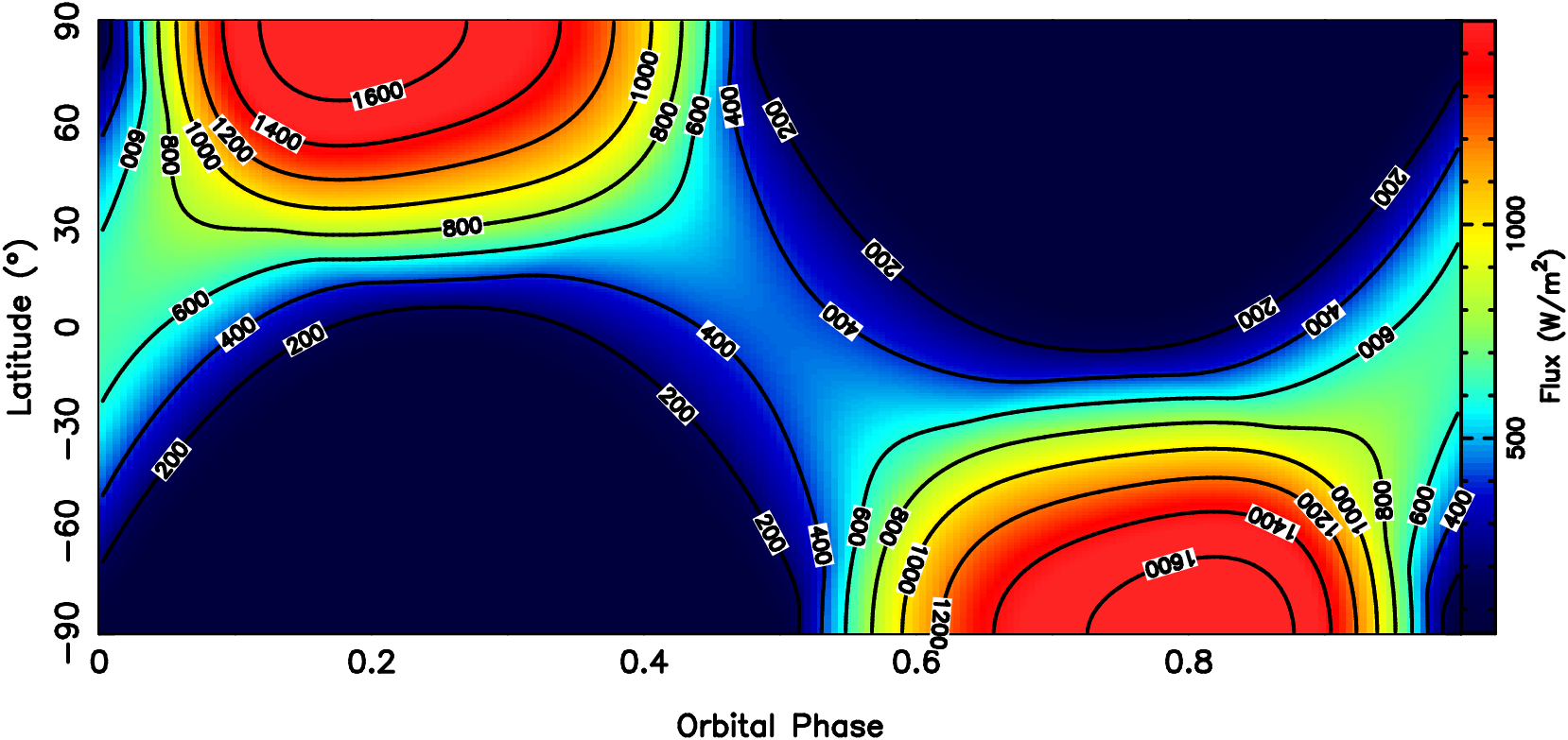}
  \end{center}
  \caption{Incident flux intensity maps of the planet GJ-163~c as a
    function of latitude and orbital phase, where phase $\phi = 0.0$
    corresponds to periastron passage of the planet. Top: $e = 0.099$,
    $\theta = 20\degr$, $\Delta \phi = 0.0$. Middle: $e = 0.099$,
    $\theta = 50\degr$, $\Delta \phi = 0.25$. Bottom: $e = 0.099$,
    $\theta = 80\degr$, $\Delta \phi = 0.25$.}
  \label{gj163fig}
\end{figure*}

Shown in Figure~\ref{gj163fig} are three incident flux maps for the
planet GJ-163~c. As with Figure~\ref{earth1}, the flux maps are a
function of latitude and orbital phase with contours of constant
flux. The phase of $\phi = 0.0$ corresponds to the planet's periastron
passage. All three panels use the known eccentricity of $e =
0.099$. The top panel assumes an obliquity of $\theta = 20\degr$ and a
phase offset between periastron and highest stellar declination in the
northern hemisphere of $\Delta \phi = 0.0$. The lower two panels
assume $\Delta \phi = 0.25$ and obliquities of $\theta = 50\degr$
(middle) and $\theta = 80\degr$ (bottom). Choosing $\Delta \phi =
0.25$ demonstrates the effect of decoupling the incident flux effects
of periastron and maximum stellar declination in a particular
hemisphere. Using the methodology of Section~\ref{aveflux}, the
eccentricity of $e \sim 0.1$ and obliquity of $\theta = 20\degr$ have
approximately equivalent effects on the seasonal variations in flux at
low latitudes, and thus supply similar driving energy for the climate
variations in those low latitude regions. The middle and bottom panels
of Figure~\ref{gj163fig} show that the obliquity becomes the dominant
source of variable energy for $\theta > 20\degr$, with a mean incident
flux of 0.98~$F_\oplus$ in the latitude range of $-30\degr > \beta >
+30\degr$ for $\theta = 50\degr$.

\subsection{K2-3}
\label{k23}

An early result from the K2 mission \citep{how14} was the discovery of
the planetary system K2-3 by \citet{cro15}. The system parameters were
subsequently refined further by the work of \citet{alm15} and
\citet{sin16}. The stellar and planetary properties shown in
Table~\ref{planets} are those from \citet{cro15}, as they provide a
self-consistent model of the system. Using the stability criteria
described in Section~\ref{stability}, we calculated the estimated
planet mass and subsequent limits on the orbital eccentricity of the
outermost planet known in the system, planet d. For a circular orbit,
the semi-major axis of planet d corresponds to the inner edge of the
OHZ and the Hill radius is 0.004~AU (see Table~\ref{planets}). By
adjusting the eccentricity of the planet, the limit of $\Delta \sim 9$
is reached at an eccentricity of $e = 0.162$ where the mutual Hill
radius for the outer two planets is $R_{H,M_p} = 0.004$~AU. Adopting
this eccentricity for the outer planet results in an orbital
architecture that is depicted in the top-right panel of
Figure~\ref{hzfig}, where planet d enters the OHZ during
apastron. Although we have selected an argument of periastron of
$\omega = 90\degr$, the value of $\omega$ has no impact on our flux
calculations and the transit duration will provide limits on the
allowed periastron values for a given eccentricity \citep{kan08}.

\begin{figure*}
  \begin{center}
    \includegraphics[angle=270,width=16.0cm]{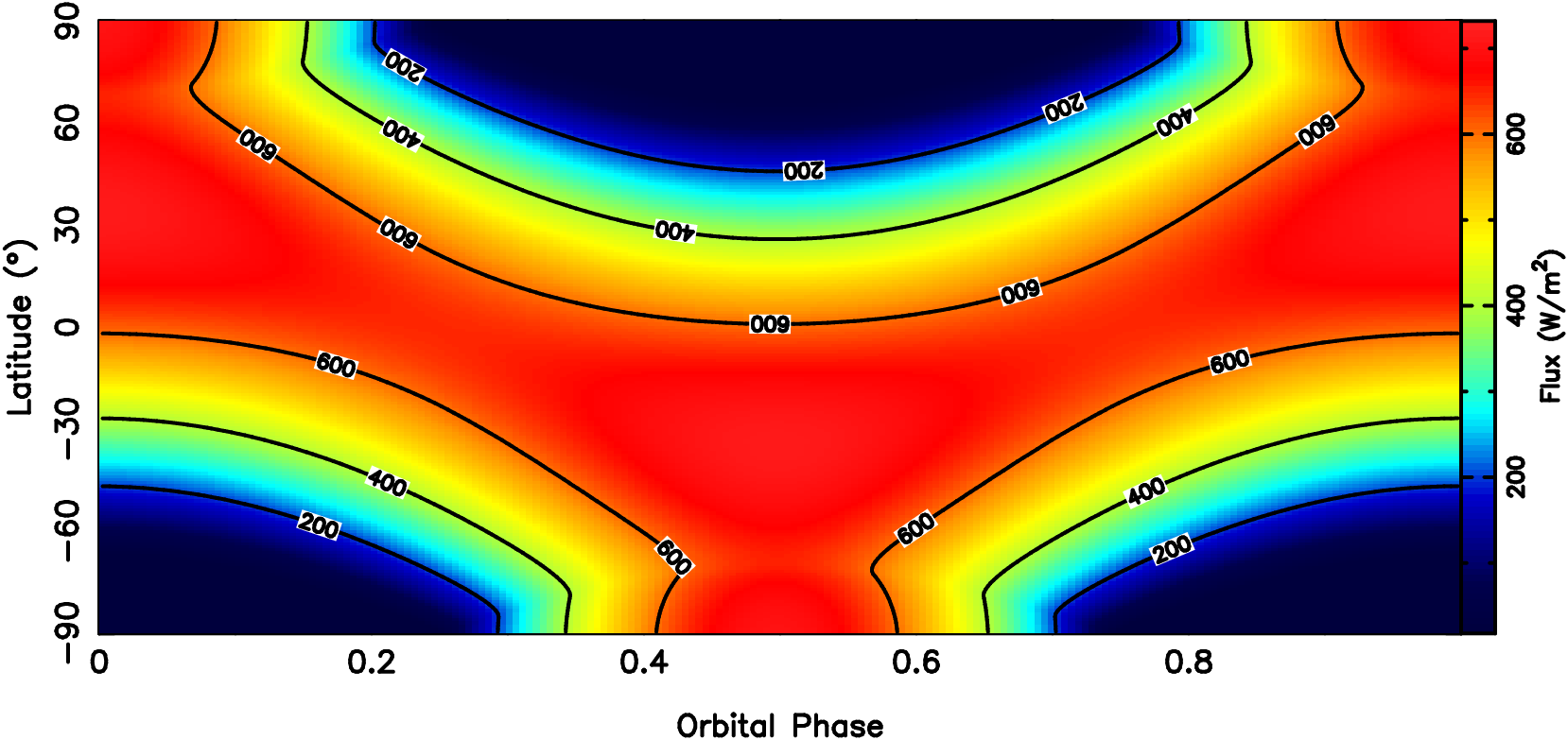} \\
    \includegraphics[angle=270,width=16.0cm]{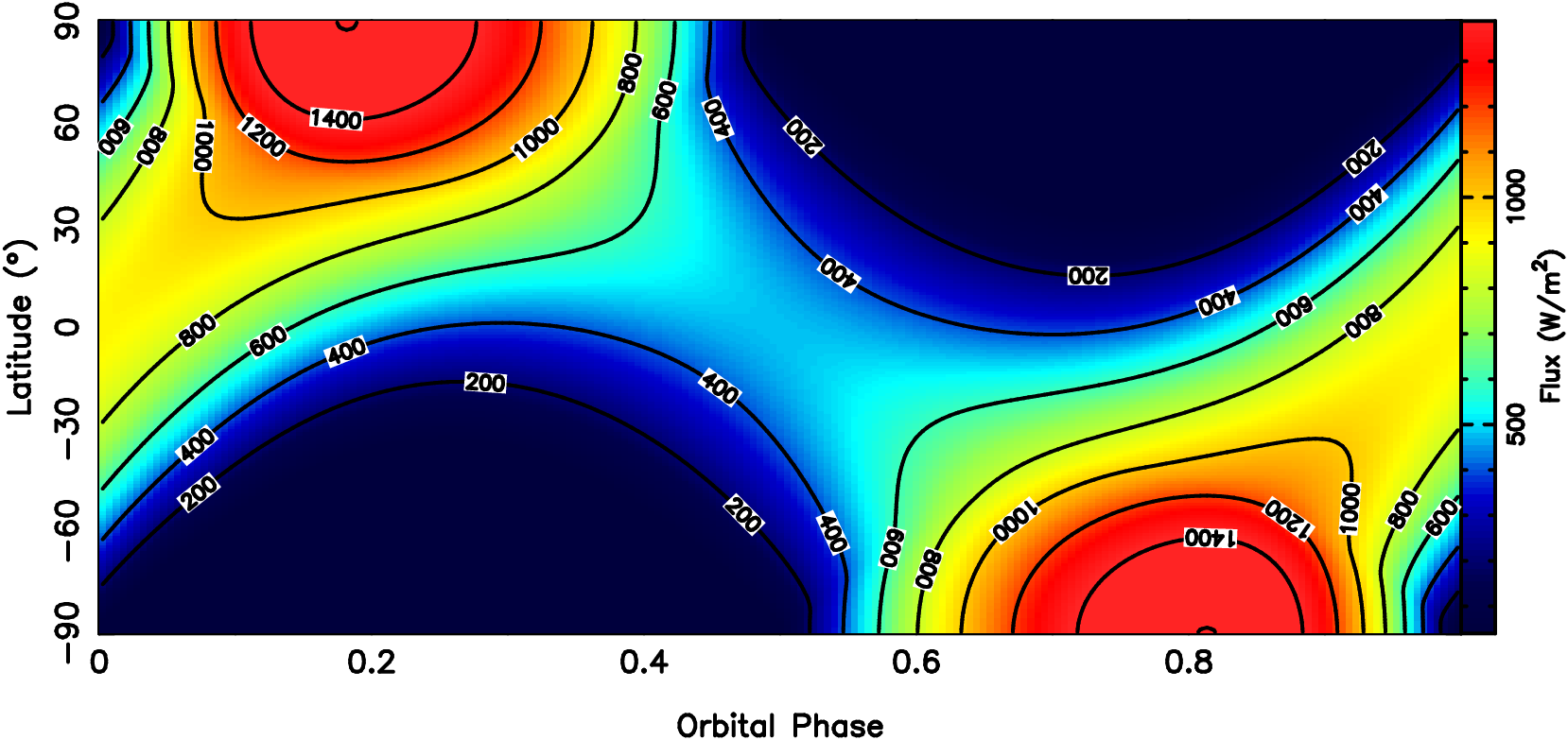} \\
    \includegraphics[angle=270,width=16.0cm]{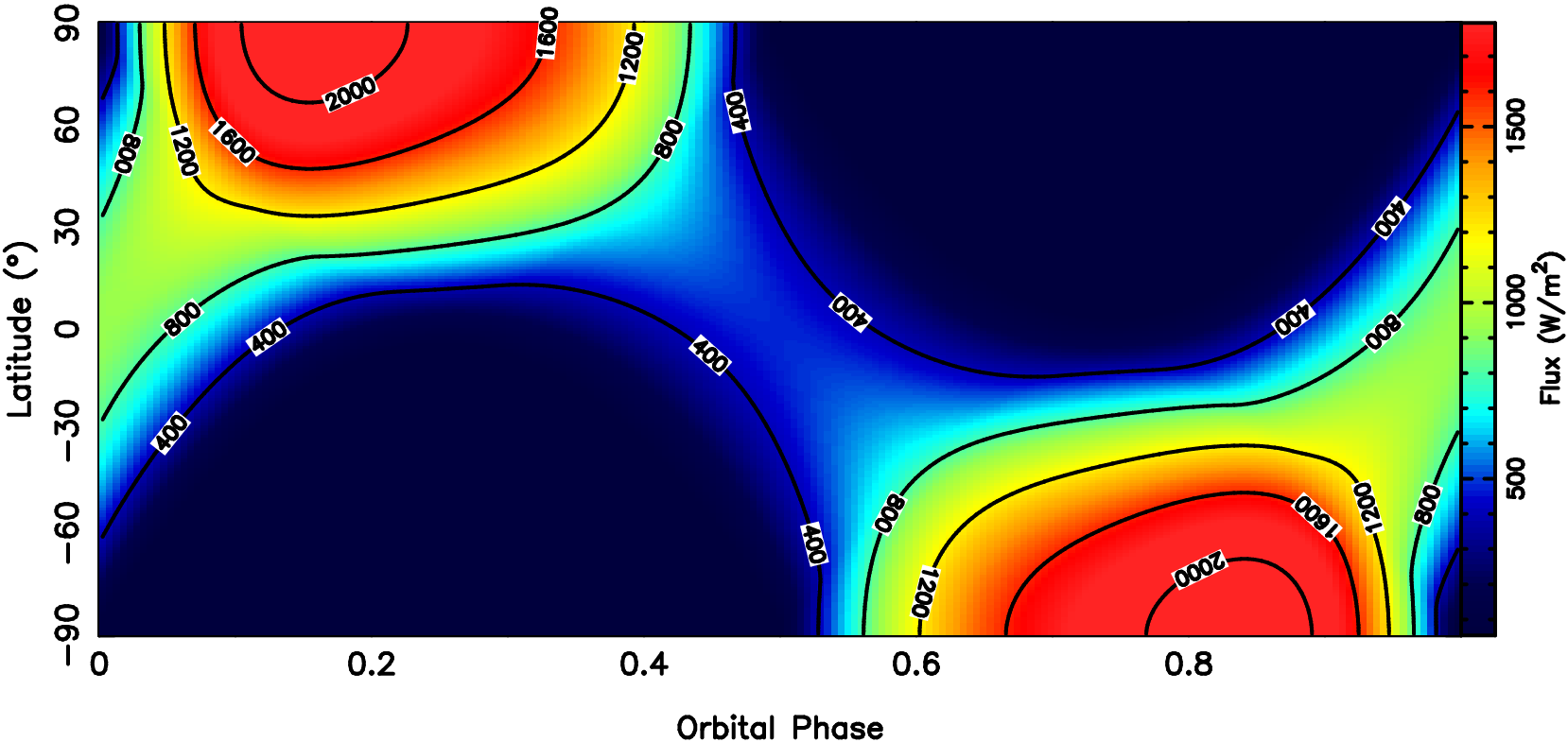}
  \end{center}
  \caption{Incident flux intensity maps of the planet K2-3~d as a
    function of latitude and orbital phase, where phase $\phi = 0.0$
    corresponds to periastron passage of the planet. Top: $e = 0.0$,
    $\theta = 20\degr$, $\Delta \phi = 0.0$. Middle: $e = 0.162$,
    $\theta = 50\degr$, $\Delta \phi = 0.25$. Bottom: $e = 0.162$,
    $\theta = 80\degr$, $\Delta \phi = 0.25$.}
  \label{k23fig}
\end{figure*}

From Figure~\ref{fluxvar2}, it can be seen that the eccentricity of $e
= 0.162$ results in an equivalent flux variation to an obliquity of
$\theta = 17\degr$ at a latitude of $\beta = 20\degr$. If planet d is
in a circular orbit, the flux received by the planet during the entire
orbit is 2055~W\,m$^{-2}$, (1.50~$F_\oplus$). For an eccentricity of
$e = 0.162$, the maximum flux received is 2925~W\,m$^{-2}$
(2.14~$F_\oplus$). Shown in Figure~\ref{k23fig} are the flux intensity
maps for K2-3~d as a function of latitude and orbital phase where,
once again, we include the diurnal effects. The top panel assumes a
circular orbit, an obliquity of $\theta = 20\degr$, and $\Delta \phi =
0.0$. The bottom two panels assume a maximum eccentricity of $e =
0.162$, $\Delta \phi = 0.25$, and obliquities of $50\degr$ and
$80\degr$ for the middle and bottom panels respectively.  The
amplitude of the flux variation effects in the top panel are below
those predicted by the maximum eccentricity and would thus result in a
more temperate climate than the eccentric cases in the bottom two
panels. The mean incident flux in the latitude range of $-30\degr >
\beta > +30\degr$ for $\theta = 50\degr$ (Figure~\ref{k23fig}, middle
panel) is 1.20~$F_\oplus$.

\begin{figure*}
  \begin{center}
    \includegraphics[angle=270,width=16.0cm]{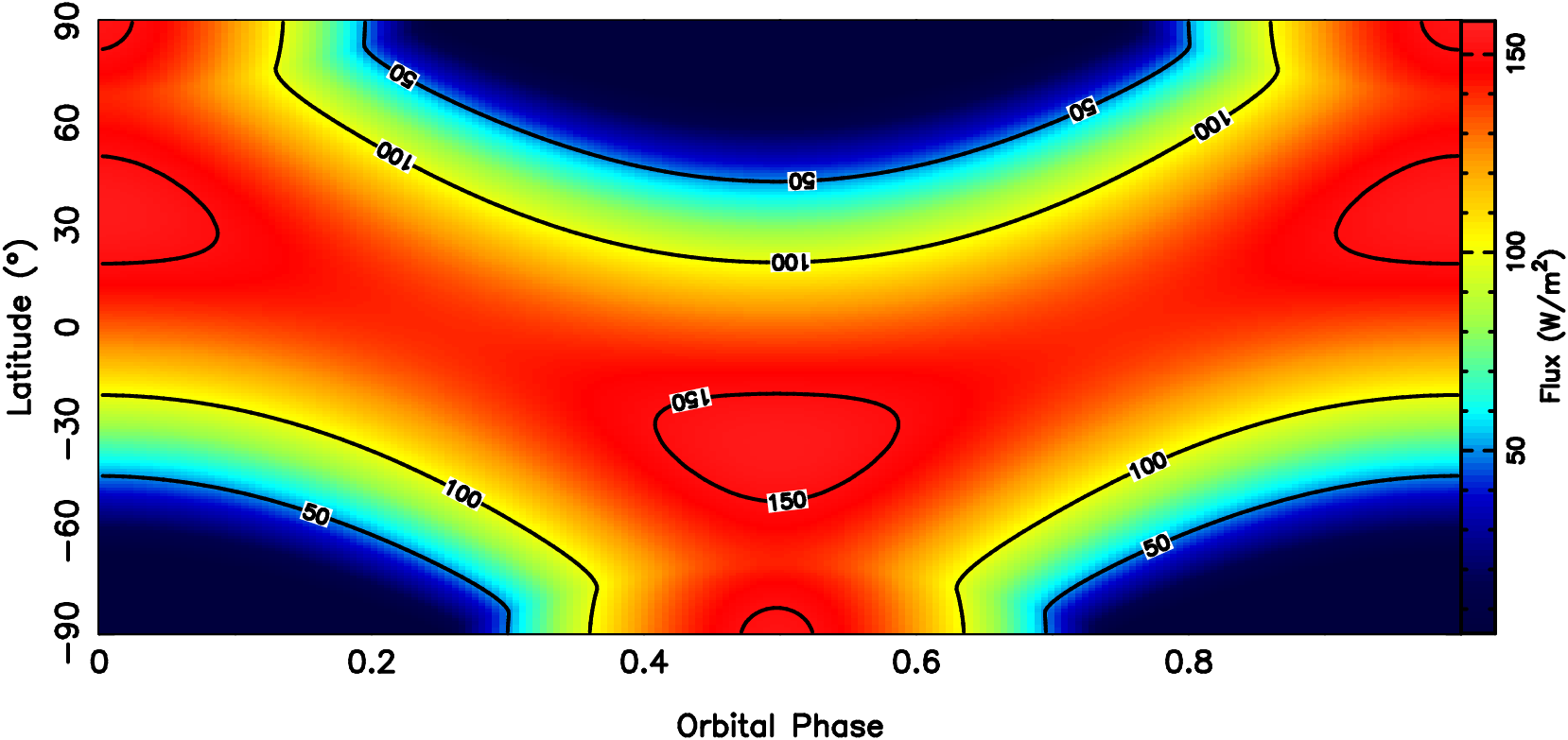} \\
    \includegraphics[angle=270,width=16.0cm]{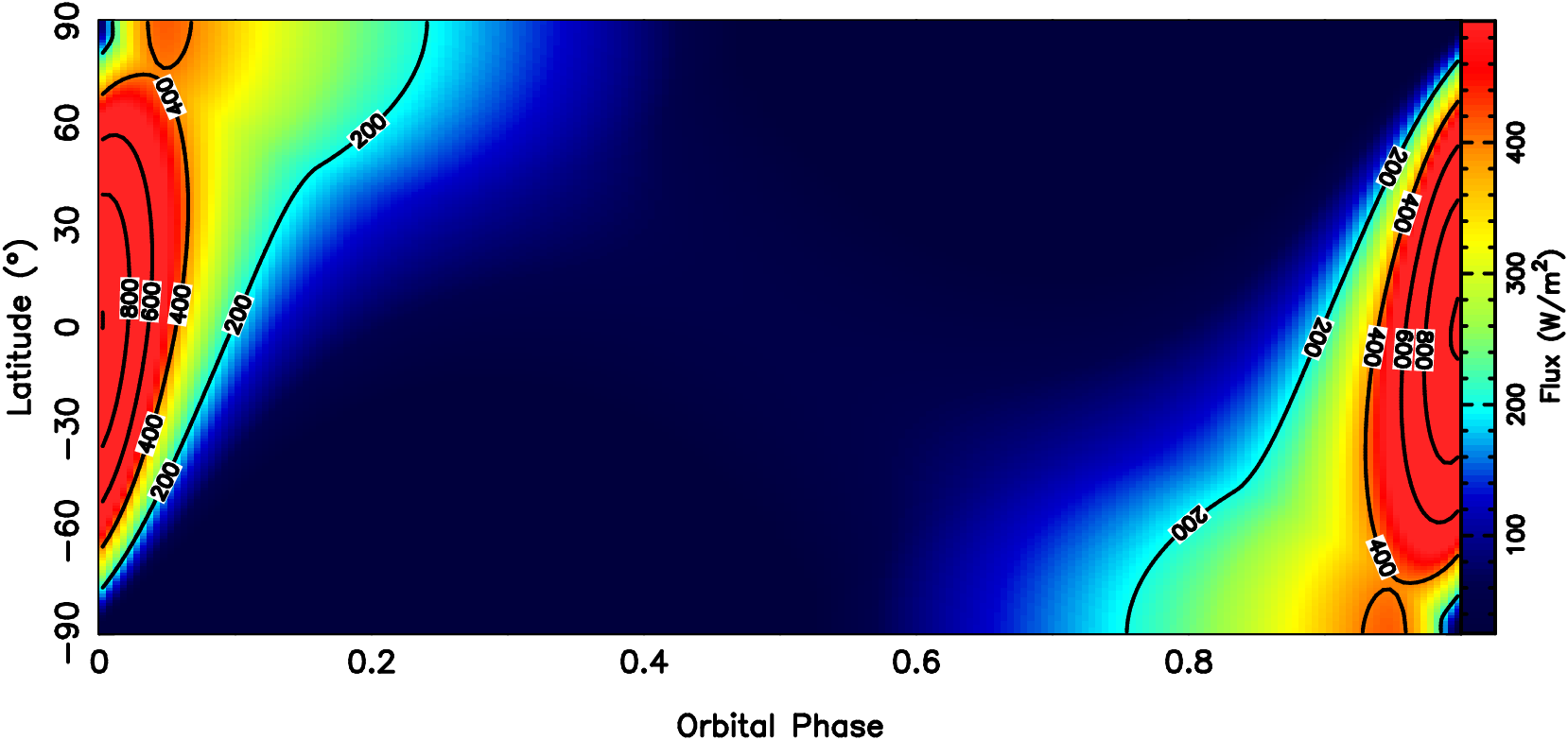} \\
    \includegraphics[angle=270,width=16.0cm]{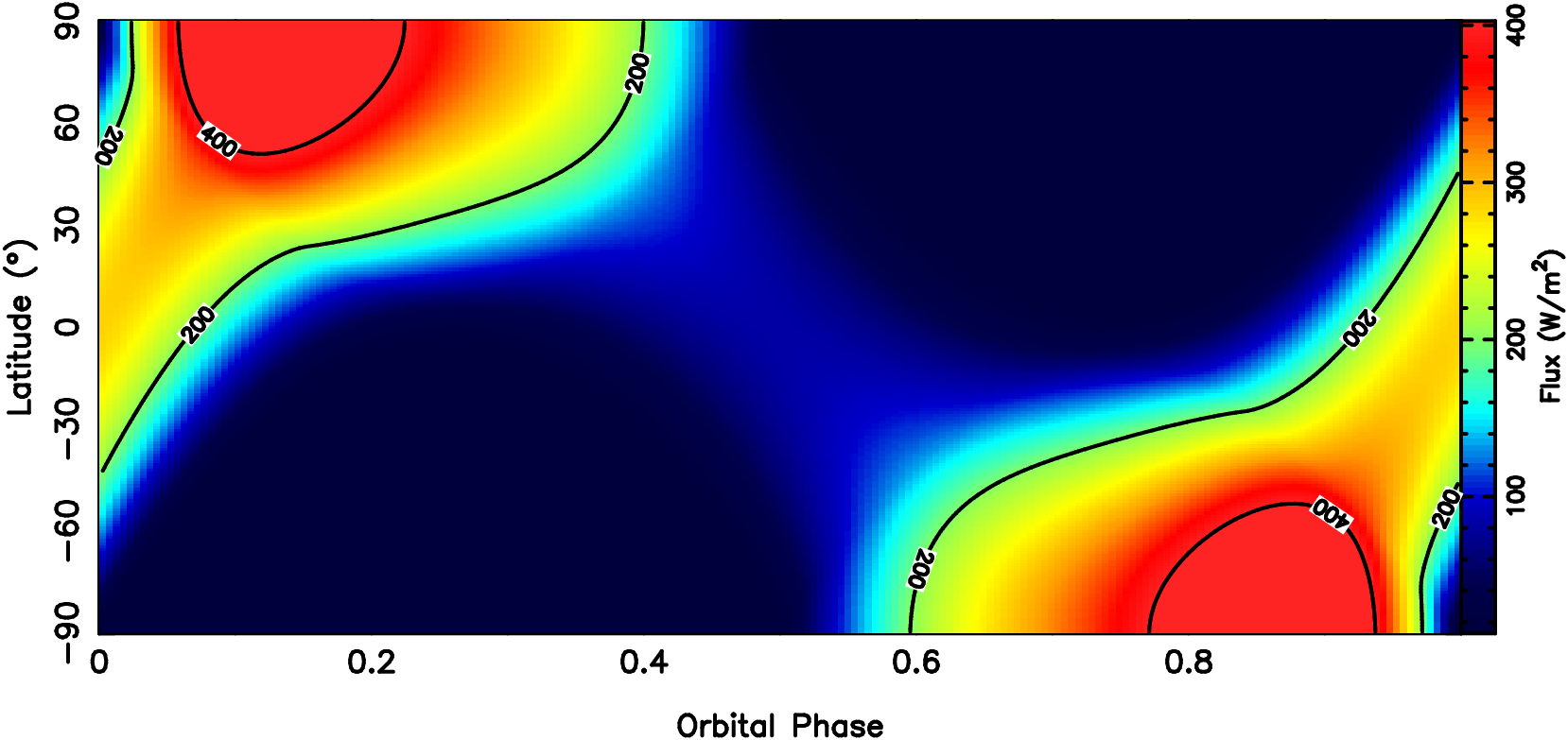}
  \end{center}
  \caption{Incident flux intensity maps of the planet Kepler-186~f as
    a function of latitude and orbital phase, where phase $\phi = 0.0$
    corresponds to periastron passage of the planet. Top: $e = 0.0$,
    $\theta = 20\degr$, $\Delta \phi = 0.0$. Middle: $e = 0.628$,
    $\theta = 50\degr$, $\Delta \phi = 0.25$. Bottom: $e = 0.3$,
    $\theta = 80\degr$, $\Delta \phi = 0.25$.}
  \label{kepler186fig}
\end{figure*}

\begin{figure*}
  \begin{center}
    \includegraphics[angle=270,width=16.0cm]{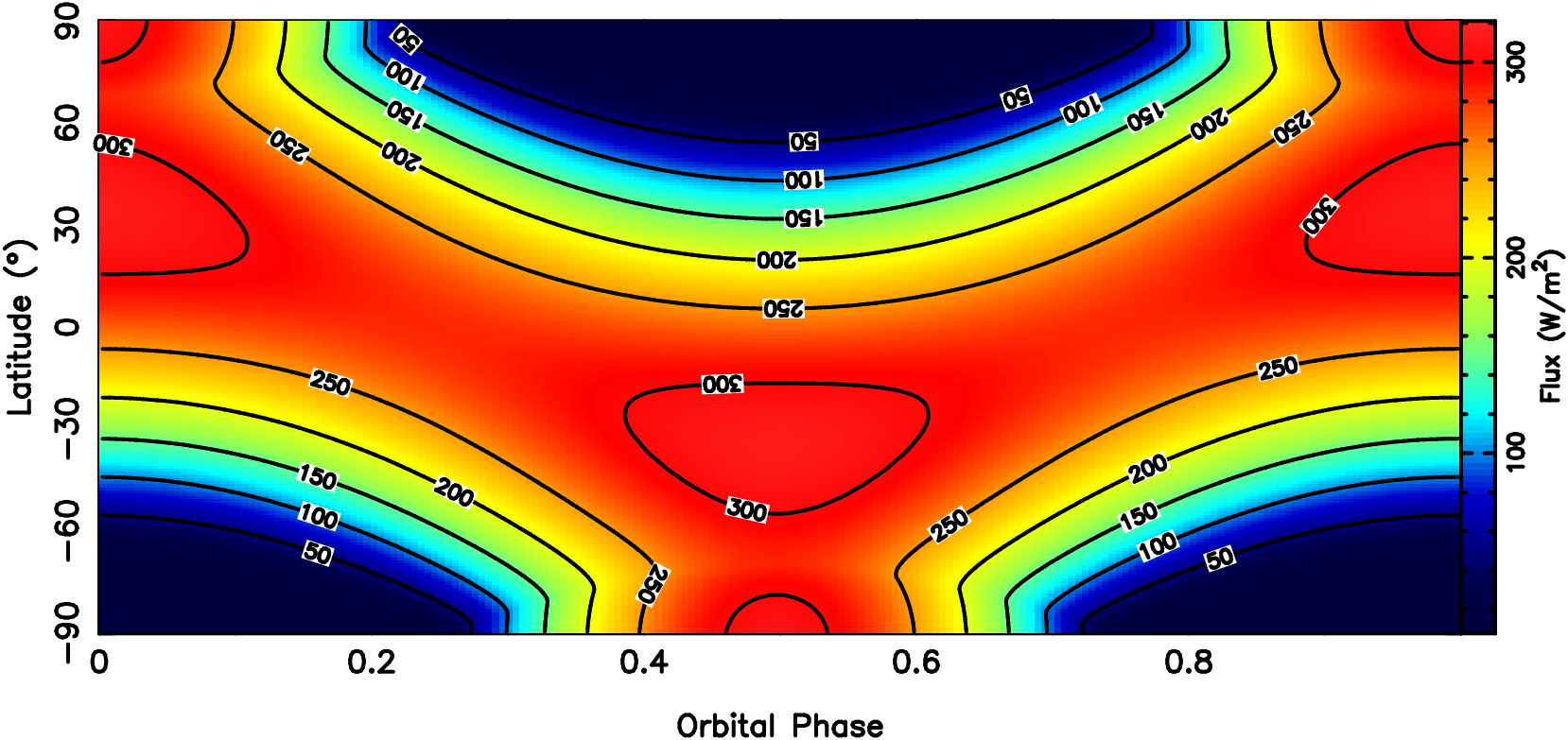} \\
    \includegraphics[angle=270,width=16.0cm]{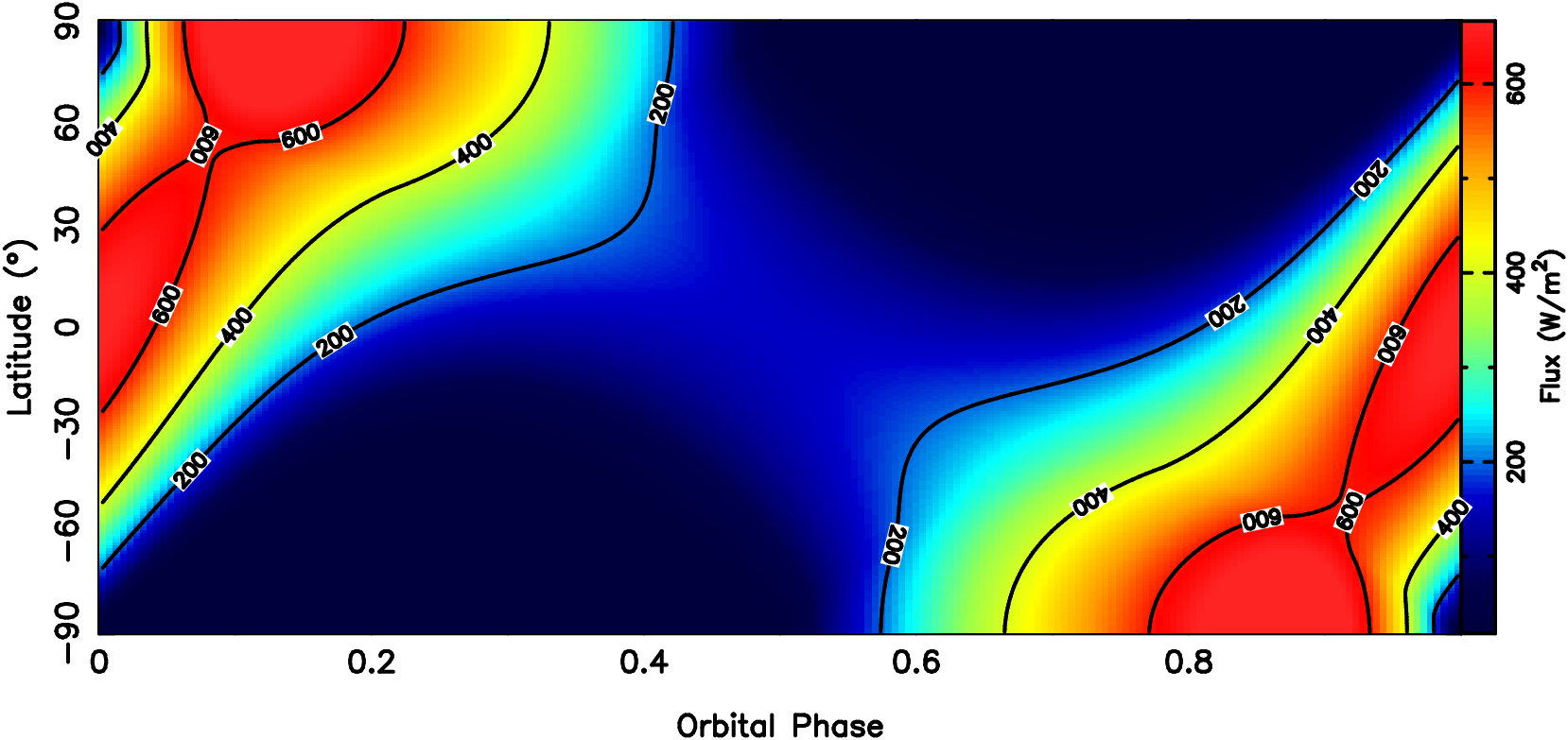} \\
    \includegraphics[angle=270,width=16.0cm]{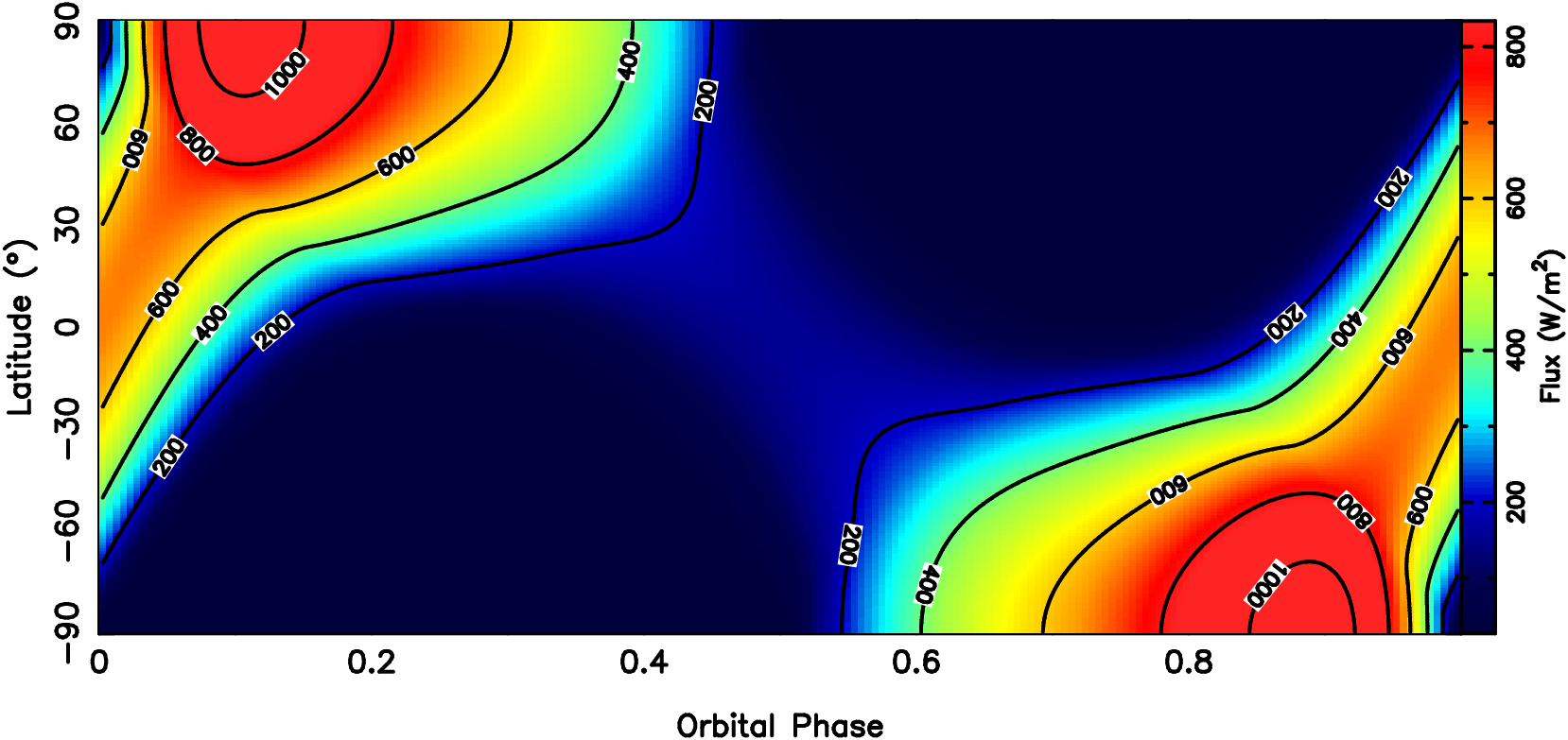}
  \end{center}
  \caption{Incident flux intensity maps of the planet Proxima
    Centauri~b as a function of latitude and orbital phase, where
    phase $\phi = 0.0$ corresponds to periastron passage of the
    planet. Top: $e = 0.0$, $\theta = 20\degr$, $\Delta \phi =
    0.0$. Middle: $e = 0.35$, $\theta = 50\degr$, $\Delta \phi =
    0.25$. Bottom: $e = 0.35$, $\theta = 80\degr$, $\Delta \phi =
    0.25$.}
  \label{proximafig}
\end{figure*}


\subsection{Kepler-186}
\label{kepler186}

The multi-planet system, Kepler-186, was confirmed by \citet{lis14}
and \citet{row14} and later confirmed to have a fifth planet by
\citet{qui14}. The new outer planet, designated Kepler-186~f, was a
particularly important discovery due to its relatively small size and
location within the HZ of the host star \citet{bol14}. Our adopted
stellar and planetary properties for the Kepler-186 system are from
\citet{qui14} and are shown in Table~\ref{planets}. Combining these
parameters with the methodology of Section~\ref{stability} results in
a maximum eccentricity of the outer planet of $e = 0.628$ where the
mutual Hill radius for the outer two planets is $R_{H,M_p} =
0.0025$~AU. Adopting this eccentricity for Kepler-186~f results in the
orbital architecture depicted in the bottom-left panel of
Figure~\ref{hzfig}. We will explore the effect of this extreme
eccentricity limit on flux variations noting that, as for K2-3~d, the
periastron argument for an eccentric orbit may be constrained from the
transit duration.

As can be seen in Figure~\ref{fluxvar1} and Figure~\ref{fluxvar2},
such an extreme eccentricity has no obliquity equivalent at latitude
$\beta = 20\degr$ regardless of diurnal effects. Such
obliquity-induced flux variations are only possible for $\theta >
60\degr$ for the maximum flux case and $\theta > 30\degr$ for the
diurnal case. Planet f is toward the outer edge of the system HZ and,
for a circular orbit, receives a maximum flux of 445~W\,m$^{-2}$
(0.33~$F_\oplus$). Adopting the extreme eccentricity of $e = 0.628$
creates a significant change in this result, with a maximum flux
during periastron passage of 3212~W\,m$^{-2}$ (2.35~$F_\oplus$). The
flux intensity maps for Kepler-186~f as a function of latitude and
orbital phase using the diurnal model are shown in
Figure~\ref{kepler186fig}. The top panel shows the flux map for the
scenario of a circular orbit, an obliquity of $\theta = 20\degr$, and
an alignment of maximum stellar declination and periastron passage
($\Delta \phi = 0.0$). The bottom two panels assume a maximum stellar
declination phase offset from periastron passage of $\Delta \phi =
0.25$. The middle panel represents the extreme eccentricity scenario
with $e = 0.628$ and an obliquity of $\theta = 50\degr$ and further
demonstrates how the eccentricity dominates the flux variations for
even relatively high obliquities. The scenario shown in the bottom
panel assumes a more moderate eccentricity of $e = 0.3$ where the
obliquity of $\theta = 80\degr$ is more readily able to drive the
seasonal flux variations.


\subsection{Proxima Centauri}
\label{proxima}

The terrestrial planet orbiting Proxima Centauri was discovered by
\citet{ang16}. This is a naturally high-value planet as it is, by
definition, the closest exoplanet to our planetary system. The value
is increased by its orbit lying within the HZ of the host star,
leading to the exploration of potentially habitable conditions and
detectable biosignatures \citep{bar17,mea17,rib16,tur16}. Although
they are not entirely ruled out, no evidence of planetary transits
have been found at this time \citep{ang16,dav16}. The orbital solution
provided by \citet{ang16} has a maximum orbital eccentricity of $e =
0.35$. \citet{kan17} utilized this eccentricity to calculate
observable signatures as a function of planet mass, and they also
performed stability simulations that exclude the presence of
additional terrestrial planets in the HZ of the system. The orbit of
the planet in relation to the system HZ is shown in the bottom-right
panel of Figure~\ref{hzfig}, where the maximum eccentricity has been
adopted.

According to Figure~\ref{fluxvar2}, the maximum eccentricity of $e =
0.35$ produces a flux variation equivalent to an obliquity of $\theta
= 44\degr$ at a latitude of $\beta = 20\degr$. For the circular orbit
scenario, the maximum flux received by the planet is 901~W\,m$^{-2}$
(0.66~$F_\oplus$), whereas the eccentric scenario results in a maximum
incident flux of 2133~W\,m$^{-2}$ (1.56~$F_\oplus$). The flux
intensity maps for Proxima Centauri b as a function of latitude and
orbital phase are shown in Figure~\ref{proximafig}. The top panel
represents the circular orbit case along with an obliquity of $\theta
= 20\degr$ and an alignment of maximum stellar declination and
periastron passage ($\Delta \phi = 0.0$). The bottom two panels of
Figure~\ref{proximafig} represent the maximum eccentricity case and
assume a maximum stellar declination phase offset from periastron
passage of $\Delta \phi = 0.25$. The middle panel shows the flux map
for an obliquity of $\theta = 50\degr$ and thus represents the case
where the flux variations match those of the eccentricity at latitude
$\beta = 20\degr$. The bottom panel assumes an obliquity of $\theta =
80\degr$. Fully constraining the eccentricity of this planet (and,
indeed, of all planets) is clearly critical for developing the needed
flux maps to determine climate cycles and potential impacts on surface
temperatures.


\section{Implications for Habitability}
\label{implications}

The construction of detailed GCMs relies heavily upon many factors,
such as the atmospheric composition, temperature-pressure profile, and
orbital properties (see references provided in
Section~\ref{intro}). With relatively few exceptions, measurements of
exoplanet parameters are currently restricted to the mass, radius, and
Keplerian orbital properties. Parameters that are inaccessible, at
least for terrestrial planets, include the planetary rotation rate and
the obliquity of the rotation axis. The influence of rotation rate on
atmospheric dynamics for HZ planets has been considered in detail
\citep{yan14,lec15}, and it has been shown that the evolution of cloud
layers at the substellar point that influence habitable surface
conditions is highly sensitive to the rotation period
\citep{kop16}. It is therefore important to include the diurnal
effects that we have incorporated into our flux map models, as
described in Section~\ref{aveflux}.

The effect of obliquity on habitable climates is substantial, such as
the possibility for HZ planets with large obliquities to experience
regular global snowball transitions \citep{spi09}. For the Earth, the
obliquity is stabilized by the Earth's moon \citep{las93b,li14},
without which the obliquity variations would likely have been much
more extreme \citep{las93a}. Additional simulations for a
retrograde-rotating Venus by \citet{bar16} indicate that obliquity
variations may have been as low as $\pm 7\degr$ over Gyr timescales,
implying that massive moons are not necessarily required for obliquity
stability. In either case, the obliquity of a particular exoplanet is
one that must float as a free parameter in the GCMs that predict
surface conditions. A direct measurement of obliquity from seasonal
variations in directly detected light will be possible from future
missions capable of such measurements. Modeling of these data using
current Earth-based observations shows that planetary rotational and
obliquity parameters may be inferred from exoplanet imaging photometry
\citep{kan15,kaw16,sch16}.

A planetary parameter that can be presently measured is the orbital
eccentricity. This parameter is most often extracted from the
Keplerian orbital solution to RV observations of a bright host star
and can also be inferred to a lesser extent from the duration of a
planetary transit \citep{bar08,bur08}. The eccentricities for most of
the Kepler HZ planets are largely unknown due to the faintness of the
host stars \citep{kan16}. In addition, variable eccentricities due to
dynamical interactions with other planets can induce Milankovitch
cycles with significantly shorter periods than those measured from the
Earth \citep{way17}. The primary purpose of the study described in
this work could then be seen as placing constraints on the variable
flux from the measurable parameter of eccentricity as a proxy for the
presently unknown obliquity of the exoplanet.


\section{Conclusions}
\label{conclusions}

Despite the rapid progress in our understanding of terrestrial
exoplanets frequency in the HZ, there are many planetary parameters
crucial to calculating habitability models that remain beyond our
reach. The seasonal variations in incident flux are driven by the
orbital eccentricity and the obliquity of the planet's rotational
angular momentum. Within the solar system, Mars is an example of a
planet where the obliquity and eccentricity play similar roles in
driving the seasonal climate variations. Of the two, eccentricity is
currently our only accessible parameter and so it is useful to
determine the limits on seasonal variations imposed by the
eccentricity that would be matched by a particular obliquity.

In this work, we have calculated the effects of eccentricity and
obliquity on incident flux as a function of latitude, and where the
flux variations are equivalent for a complete orbital cycle. The two
effects largely differ in the Keplerian nature of the eccentricity
variations as opposed to the sinusoidal changes in obliquity-induced
flux at a given latitude. We selected four case studies of terrestrial
planets in the HZ of their host stars where the eccentricity is either
measured or we were able to calculate a maximum dynamical
eccentricity. These case studies demonstrate where extreme
eccentricities and obliquities can dominate the incident flux map and
is particularly important in demonstrating the contrast to either the
zero eccentricity and/or zero obliquity models. This in turn
establishes the importance of constraining eccentricity, as even a
relatively small eccentricity ($e \sim 0.2$) can have a large
influence on the flux map and climate cycles. Until such time as
direct measurements of obliquity can be made, the models presented
here will find their utility in constraining obliquity for a given
eccentricity and flux map of the planet.


\section*{Acknowledgements}

The authors would like to thank Fred Adams, Colin Chandler, and Ravi
Kopparapu for useful discussions regarding this work. The authors
would also like to thank the anonymous referee, whose comments greatly
improved the quality of the paper. This research has made use of the
Habitable Zone Gallery at hzgallery.org. This research has also made
use of the NASA Exoplanet Archive, which is operated by the California
Institute of Technology, under contract with the National Aeronautics
and Space Administration under the Exoplanet Exploration Program. The
results reported herein benefited from collaborations and/or
information exchange within NASA's Nexus for Exoplanet System Science
(NExSS) research coordination network sponsored by NASA's Science
Mission Directorate.



\begin{thebibliography}{}

\bibitem[Akeson et al.(2013)]{ake13} Akeson, R.A., Chen, X., Ciardi,
  D., et al. 2013, PASP, 125, 989
\bibitem[Almenara et al.(2015)]{alm15} Almenara, J.M.,
  Astudillo-Defru, N., Bonfils, X., et al. 2015, A\&A, 581, L7
\bibitem[Anglada-Escud\'e et al.(2016)]{ang16} Anglada-Escud\'e, G.,
  Amado, P.J., Barnes, J., et al. 2016, Nature, 536, 437
\bibitem[Armstrong et al.(2014)]{arm14} Armstrong, J.C., Barnes, R.,
  Domagal-Goldman, S., et al. 2014, AsBio, 14, 277
\bibitem[Barnes et al.(2008)]{bar08} Barnes, R., Raymond, S.N.,
  Jackson, B., Greenberg, R. 2008, AsBio, 8, 557
\bibitem[Barnes et al.(2009)]{bar09} Barnes, R., Jackson, B.,
  Greenberg, R., Raymond, S.N. 2009, ApJ, 700, L30
\bibitem[Barnes et al.(2013)]{bar13} Barnes, R., Mullins, K.,
  Goldblatt, C., et al. 2013, AsBio, 13, 225
\bibitem[Barnes et al.(2016)]{bar16} Barnes, J.W., Quarles, B.,
  Lissauer, J.J., Chambers, J., Hedman, M.M. 2016, AsBio, 16, 487
\bibitem[Barnes et al.(2017)]{bar17} Barnes, R., Deitrick, R., Luger,
  R., et al. 2017, AsBio, submitted (arXiv:1608.06919)
\bibitem[Bolmont et al.(2014)]{bol14} Bolmont, E., Raymond, S.N., von
  Paris, P., et al. 2014, ApJ, 793, 3
\bibitem[Bolmont et al.(2016)]{bol16} Bolmont, E., Libert, A.-S.,
  Leconte, J., Selsis, F. 2016, A\&A, 591, A106
\bibitem[Bonfils et al.(2013)]{bon13} Bonfils, X., Lo Curto, G.,
  Correia, A.C.M., et al. 2013, A\&A, 556, A110
\bibitem[Burke(2008)]{bur08} Burke, C.J. 2008, ApJ, 679, 1566
\bibitem[Crossfield et al.(2015)]{cro15} Crossfield, I.J.M., Petigura,
  E., Schlieder, J.E., et al. ApJ, 804, 10
\bibitem[Davenport et al.(2016)]{dav16} Davenport, J.R.A., Kipping,
  D.M., Sasselov, D., Matthews, J.M., Cameron, C. 2016, ApJ, 829, L31
\bibitem[Dressing et al.(2010)]{dre10} Dressing, C.D., Spiegel, D.S.,
  Scharf, C.A., Menou, K., Raymond, S.N. 2010, ApJ, 721, 1295
\bibitem[Driscoll \& Barnes(2015)]{dri15} Driscoll, P.E., Barnes,
  R. 2015, AsBio, 15, 739
\bibitem[Fressin et al.(2013)]{fre13} Fressin, F., Torres, G.,
  Charbonneau, D., et al. 2013, ApJ, 766, 81
\bibitem[Gladman(1993)]{gla93} Gladman, B. 1993, Icarus, 106, 247
\bibitem[Howard(2013)]{how13} Howard, A.W. 2013, Science, 340, 572
\bibitem[Howell et al.(2014)]{how14} Howell, S.B., Sobeck, C., Haas,
  M., et al. 2014, PASP, 126, 398
\bibitem[Kane \& von Braun(2008)]{kan08} Kane, S.R., von Braun,
  K. 2008, ApJ, 689, 492
\bibitem[Kane \& Gelino(2012a)]{kan12a} Kane, S.R., Gelino,
  D.M. 2012a, PASP, 124, 323
\bibitem[Kane \& Gelino(2012b)]{kan12b} Kane, S.R., Gelino,
  D.M. 2012b, AsBio, 12, 940
\bibitem[Kane et al.(2012)]{kan12c} Kane, S.R., Ciardi, D.R., Gelino,
  D.M., von Braun, K. 2012, MNRAS, 425, 757
\bibitem[Kane et al.(2015)]{kan15} Kane, S.R., Domagal-Goldman, S.D.,
  Herman, J.R., Robinson, T.D., Stine, A.R. 2015, Proceedings of the
  Comparative Climates of Terrestrial Planets II (arXiv:1511.03779)
\bibitem[Kane et al.(2017)]{kan16} Kane, S.R., Hill, M.L., Kasting,
  J.F., et al. 2016, ApJ, 830, 1
\bibitem[Kane et al.(2017)]{kan17} Kane, S.R., Gelino, D.M., Turnbull,
  M.C. 2017, AJ, 153, 52
\bibitem[Kasting et al.(1993)]{kas93} Kasting, J.F., Whitmire, D.P.,
  Reynolds, R.T. 1993, Icarus, 101, 108
\bibitem[Kaspi \& Showman(2015)]{kas15} Kaspi, Y., Showman, A.P. 2015,
  ApJ, 804, 60
\bibitem[Kawahara(2016)]{kaw16} Kawahara, H. 2016, ApJ, 822, 112
\bibitem[Kopparapu et al.(2013)]{kop13} Kopparapu, R.K., Ramirez, R.,
  Kasting, J.F., et al. 2013, ApJ, 765, 131
\bibitem[Kopparapu et al.(2014)]{kop14} Kopparapu, R.K., Ramirez,
  R.M., SchottelKotte, J., et al. 2014, ApJ, 787, L29
\bibitem[Kopparapu et al.(2016)]{kop16} Kopparapu, R.K., Wolf, E.T.,
  Haqq-Misra, J., et al. 2016, ApJ, 819, 84
\bibitem[Laskar(1986)]{las86} Laskar, J. 1986, A\&A, 157, 59
\bibitem[Laskar \& Robutel(1993)]{las93a} Laskar, J., Robutel,
  P. 1993, Nature, 361, 608
\bibitem[Laskar et al.(1993)]{las93b} Laskar, J., Joutel, F., Robutel,
  P. 1993, Nature, 361, 615
\bibitem[Leconte et al.(2013)]{lec13} Leconte, J., Forget, F.,
  Charnay, B., et al. 2013, A\&A, 554, A69
\bibitem[Leconte et al.(2015)]{lec15} Leconte, J., Wu, H., Menou, K.,
  Murray, N. 2015, Science, 10, 1126
\bibitem[Li \& Batygin(2014)]{li14} Li, G., Batygin, K. 2014, ApJ,
  790, 69
\bibitem[Linsenmeier et al.(2015)]{lin15} Linsenmeier, M., Pascale,
  S., Lucarini, V. 2015, P\&SS, 105, 43
\bibitem[Lissauer et al.(2014)]{lis14} Lissauer, J.J., Marcy, G.W.,
  Bryson, S.T., et al. 2014, ApJ, 784, 44
\bibitem[Meadows et al.(2017)]{mea17} Meadows, V.S., Arney, G.N.,
  Schwieterman, E.W., et al. 2017, Astrobiology, submitted
  (arXiv:1608.08620)
\bibitem[Petigura et al.(2013)]{pet13} Petigura, E.A., Marcy, G.W.,
  Howard, A.W. 2013, ApJ, 770, 69
\bibitem[Quintana et al.(2014)]{qui14} Quintana, E.V., Barclay, T.,
  Raymond, S.N., et al. 2014, Science, 344, 277
\bibitem[Ribas et al.(2016)]{rib16} Ribas, I., Bolmont, E., Selsis,
  F., et al. 2016, A\&A, 596, A111
\bibitem[Ricker et al.(2015)]{ric15} Ricker, G.R., Winn, J.N.,
  Vanderspek, R., et al. 2015, JATIS, 1, 014003
\bibitem[Rowe et al.(2014)]{row14} Rowe, J.F., Bryson, S.T., Marcy,
  G.W., et al. 2014, ApJ, 784, 45
\bibitem[Schwartz et al.(2016)]{sch16} Schwartz, J.C., Sekowski, C.,
  Haggard, H.M., Pall\'e, E., Cowan, N.B. 2016, MNRAS, 457, 926
\bibitem[Sinukoff et al.(2016)]{sin16} Sinukoff, E., Howard, A.W.,
  Petigura, E.A., et al. 2016, ApJ, 827, 78
\bibitem[Smith \& Lissauer(2009)]{smi09} Smith, A.W., Lissauer,
  J.J. 2009, Icarus, 201, 381
\bibitem[Spiegel et al.(2016)]{spi09} Spiegel, D.S., Menou, K.,
  Scharf, C.A. 2009, ApJ, 691, 596
\bibitem[Sullivan et al.(2015)]{sul15} Sullivan, P.W., Winn, J.N.,
  Berta-Thompson, Z.K., et al. 2015, ApJ, 809, 77
\bibitem[Tuomi \& Anglada-Escud\'e(2013)]{tuo13} Tuomi, M.,
  Anglada-Escud\'e, G. 2013, A\&A, 556, A111
\bibitem[Turbet et al.(2016)]{tur16} Turbet, M., Leconte, J., Selsis,
  F., et al. 2016, A\&A, 596, A112
\bibitem[Van Eylen \& Albrecht(2015)]{van15} Van Eylen, V., Albrecht,
  S. 2015, ApJ, 808, 126
\bibitem[Way \& Georgakarakos(2017)]{way17} Way, M.J., Georgakarakos,
  N. 2017, ApJ, 835, L1
\bibitem[Weiss \& Marcy(2014)]{wei14} Weiss, L.M., Marcy, G.W. 2014,
  ApJ, 783, L6
\bibitem[Williams \& Kasting(1997)]{wil97} Williams, D.M., Kasting,
  J.F. 1997, Icarus, 129, 254
\bibitem[Williams \& Pollard(2002)]{wil02} Williams, D.M., Pollard,
  D. 2002, IJAsB, 1, 61
\bibitem[Williams \& Pollard(2003)]{wil03} Williams, D.M., Pollard,
  D. 2003, IJAsB, 2, 1
\bibitem[Wolf \& Toon(2013)]{wol13} Wolf, E., Toon, O.B. 2013,
  Astrobiology, 13, 656
\bibitem[Wolf \& Toon(2014)]{wol14} Wolf, E., Toon, O.B. 2014,
  Astrobiology, 14, 241
\bibitem[Wordsworth et al.(2010)]{wor10} Wordsworth, R., Forget, F.,
  Selsis, F., et al. 2010, A\&A 522, A22
\bibitem[Wordsworth et al.(2011)]{wor11} Wordsworth, R., Forget, F.,
  Selsis, F., et al. 2011, ApJ, 733, L48
\bibitem[Yang et al.(2013)]{yan13} Yang, J., Cowan, N.B., Abbot,
  D.S. 2013, ApJ, 771, L45
\bibitem[Yang et al.(2014)]{yan14} Yang, J., Bou\'e, G., Fabrycky, D.,
  Abbot, D.S. 2014, ApJ, 787, L2

\end{thebibliography}
\end{document}